\documentclass[prd,aps,floats,twocolumn]{revtex4}
\usepackage{commath}
\pdfoutput=1
\usepackage[usenames, dvipsnames]{color}
 \usepackage{graphicx, array}
\usepackage{graphics,epsfig}
 \usepackage{amssymb}
 \usepackage{amsmath}
 \usepackage{ulem}
 \usepackage{multirow}
  \usepackage{subfigure}
\usepackage{bm}
\usepackage{txfonts}
\usepackage{cancel}
\newcolumntype{C}[1]{>{\centering\arraybackslash}m{#1}}

\def\mnras{MNRAS}
\def\jcap{JCAP}
 \def\be{\begin{equation}}
\def\ee{\end{equation}}
 \def\bi{\begin{itemize}}
 \def\ei{\end{itemize}}
  \def\ben{\begin{enumerate}}
\def\een{\end{enumerate}}
  \def\bt{\begin{tabular}}
\def\et{\end{tabular}}
\def\bc{\begin{center}}
\def\ec{\end{center}}

\def\bea{\begin{eqnarray}}
\def\eea{\end{eqnarray}}

\def\RR{\mathcal{R}}
\def\ba{\begin{eqnarray}}
\def\ea{\end{eqnarray}}

\begin{document}

\input{epsf}

\title{Efficient simulations of large scale structure in modified gravity cosmologies with comoving Lagrangian acceleration}
\author {Georgios Valogiannis and Rachel Bean.}
\affiliation{Department of Astronomy, Cornell University, Ithaca, NY 14853, USA.}
\label{firstpage}

\begin{abstract}

We implement an adaptation of the COLA approach, a hybrid scheme that combines Lagrangian perturbation theory with an N-body approach, to model non-linear collapse in chameleon and symmetron modified gravity models. Gravitational screening is modeled effectively through the attachment of a suppression factor to the linearized Klein-Gordon equations.  

The adapted COLA approach is benchmarked, with respect to an N-body code both for the $\Lambda$CDM scenario and for the modified gravity theories. It is found to perform well in the estimation of the  dark matter power spectra, with consistency of 1\%  to $k\sim2.5$ h/Mpc. Redshift space distortions  are shown to be effectively modeled through a Lorentzian parameterization with a velocity dispersion fit to the data. We find that COLA performs less well in predicting the halo mass functions, but has consistency, within $1\sigma$ uncertainties of our simulations, in the relative changes to the mass function induced by the modified gravity models relative to $\Lambda$CDM.

The results demonstrate that  COLA, proposed to enable accurate and efficient, non-linear predictions for $\Lambda$CDM, can be effectively applied to a  wider set of cosmological scenarios, with intriguing properties, for which clustering behavior needs to be understood for upcoming surveys such as LSST, DESI, Euclid and WFIRST.
 
\end{abstract}

\maketitle

\section{Introduction}
\label{sec:intro}
The nature of the unknown mechanism responsible for the accelerated expansion of the universe, as measured by Type 1a supernovae \cite{Perlmutter:1998np, Riess:2004nr}, baryon acoustic oscillations (BAO) in galaxy clustering \cite{Eisenstein:2005su,Percival:2007yw,Percival:2009xn,Kazin:2014qga}, and the Cosmic Microwave Background (CMB) \cite{Spergel:2003cb,Ade:2013zuv, Ade:2015xua}, commonly labeled as ``Dark Energy", is one of the most challenging, open questions in modern cosmology. Assuming that Einstein's General Relativity (GR) is the correct framework to describe gravity at large scales, the recent accelerative phase can be driven either by a cosmological constant term $\Lambda$ with negative pressure, or by introducing a scalar field called quintessence \cite{wetterich1988cosmology, PhysRevD.37.3406, Copeland:2006wr}. The necessary value of $\Lambda$ to account for the observed acceleration rate is extremely small, however, when compared to the predictions from high energy physics, and as a result it has to be fine-tuned \cite{Weinberg:1988cp}. Such an unattractive feature, together with the need to explore all other alternatives, has motivated the development of theories in which GR breaks down at large scales \cite{Carroll:2003wy, Capozziello:2003tk, Nojiri:2006ri}, the so called modified gravity (MG) theories. The existence of MG theories with massive scalar fields can reproduce the recent accelerative phase, however, the fifth forces that arise as a result of their coupling to matter would, in principle, cause large deviations from the tight experimental constraints of GR in the solar system \cite{Will:2005va}.

As a consequence, MG models can only be viable if they reduce to the successful GR phenomenology in the local dense environments (eg. Earth, Solar System) through a restoring screening mechanism \cite{Khoury:2010xi,Khoury:2013tda}. Based on the qualitative features of the screening mechanism they exhibit, such schemes are commonly classified in various broad classes: the ``chameleons'' \cite{PhysRevLett.93.171104, PhysRevD.69.044026}, where the scalar fields become massive and decouple in regions of high Newtonian potential, the kinetic/``$k$-Mouflage" models \cite{Babichev:2009ee,Dvali:2010jz}, in which the deviations are screened when fifth forces exceed some critical model dependent value and the Vainshtein mechanism \cite{VAINSHTEIN1972393}, that reproduces GR when large derivatives of the fifth forces are experienced. Similar, in terms of phenomenology, with the chameleons are the symmetrons \cite{PhysRevLett.104.231301,Olive:2007aj}, which exhibit the additional property of a vanishing coupling in dense regions through symmetry restoration.  

This rich spectrum of MG models are theoretically viable and offer observational consequences that are potentially distinguishable from $\Lambda$CDM through a variety of astrophysical characteristics.  
A number of spectroscopic and photometric Large Scale Structure (LSS) surveys both currently underway, e.g. Dark Energy Survey (DES)  \citep{Abbott:2005b}, HyperSuprimeCam (HSC) \footnote{http://sumire.ipmu.jp/en/}, and eBOSS \citep{Comparat:2012hz}, and coming online in the coming decade, e.g. DESI \cite{Levi:2013gra}, PSF, LSST \cite{Abell:2009aa}, Euclid \cite{Laureijs:2011gra} and WFIRST  \cite{Spergel:2013tha}, will probe the properties of gravity with remarkable precision in both the linear and non-linear regimes, using galaxy clustering, cluster counts, gravitational lensing and peculiar velocities. They offer an unprecedented opportunity to test the landscape of modified gravity theories observationally with respect to the simplest, $\Lambda$CDM scenario. As a result, simulating the structure formation in the the linear, mildly non-linear and non-linear regimes is necessary for both $\Lambda$CDM and all the alternatives. 

A variety of analytical, semi-analytical and numerical approaches have been used to study $\Lambda$CDM and dark energy scenarios in the non-linear regime. Lagrangian perturbative techniques up to first \cite{Zeldovich:1969sb,1980lssu.book.....P} or second order \cite{Bouchet:1994xp}, have been shown to produce accurate results for $\Lambda$CDM in the linear and mildly non-linear scales without having to perform a complete numerical treatment of structure formation. They fail to achieve the desired accuracy, however, at smaller, non-linear scales for which a full N-body simulation is required. In light of the computational resources necessary for  N-body simulations, and given the successes of Lagrangian Perturbation Theory (LPT), hybrid schemes have been proposed, with the aim of combining the strengths of both approaches. In this paper we focus on the Comoving Lagrangian Acceleration (COLA) hybridization scheme \cite{Tassev:2013pn}. By evolving the large scales analytically using LPT and the small scales exactly with a full N-body treatment, the COLA method manages to produce accurate results deep in the mildly non-linear regime with only a few number of time steps, making it possible to produce fast results in exchange for some accuracy. 

In modified gravity simulations, the need to accurately capture the effects of the fifth forces and the screening mechanism adds a new layer of complexity. For an exact description, one needs to solve the full Klein-Gordon equation, whose non-linearities render the procedure both challenging and computationally expensive. It is natural consequently to investigate whether an inexpensive, approximate scheme can be used instead. A linear treatment of the perturbation equations, together with the linearized Klein-Gordon equation it produces may seem efficient at first, but a more careful examination shows \cite{2012JCAP...10..002B,Brax:2013mua} that it fails to incorporate the non-linear screening effects and gives poor results. Effective approaches \cite{Winther:2014cia} have managed to implement screening successfully, however, following a phenomenological path. An ineffective but computationally fast linearized scheme, can be combined with the attachment of a screening factor for a spherically symmetric configuration, to speed up MG simulations without the sacrifice of much accuracy. 

Given the success of Lagrangian approaches in $\Lambda$CDM simulations and the need to develop effecient, but representative, realizations of the LSS in different cosmological scenarios, it is natural to see alternative routes in MG models. The benefits of LPT have already been discussed in the context of generating initial conditions, appropriate for coupled scalar field cosmologies \cite{Li:2010re} or MG models \cite{Valkenburg:2015dsa} . In this paper, we study the effectiveness of the COLA hybrid scheme, in which the linear scales are evolved exactly using LPT and the non-linear ones using N-body simulations,  for MG scenarios. As far as the N-body component is concerned, the fifth force calculation lies in the solution of the linearized KG equation and an approximate screening implementation through the thin shell factor for a dense sphere, similar to \cite{Winther:2014cia}. In chameleon-type (and symmetron) models, a scalar field acquires a very large mass within a massive object and consequently decouples due to the Yukawa suppression, so essentially only a fraction of the total mass (thin shell) contributes to the fifth force. 

\par The layout of the paper is as follows: in Sec. \ref{sec:Formalism} we first review the MG models studied and the non-linear approaches used in the analysis.
In Sec. \ref{sec:Analysis/Results} we present our results, assessing the performance for the scheme to predict a number of LSS observables, including the matter power spectrum, the redshift space distortions, and halo mass function, before summarizing the findings and discussing implications for future work in Sec. \ref{sec:conclusions}.

\section{Formalism}
\label{sec:Formalism}

\subsection{Modified gravity and screening models}
\label{sec:Modified gravity and screening models}

A wide class of viable scalar-tensor theories have been shown to be described by a Horndeski Lagrangian \cite{Horndeski1974,PhysRevD.84.064039}. Using a general single scalar field Lagrangian,  in the Einstein frame, written in terms of a scalar field $\phi$ and its derivatives,
\begin{equation}
\mathcal{L}=\frac{M_{Pl}^2}{2}R+\mathcal{L}(\phi,\partial_{\mu} \phi,\partial_{\mu}\partial^{\mu}\phi)+\mathcal{L}_m(e^{2\beta(\phi)\phi/M_{pl}}g_{\mu\nu},\psi_m), 
\end{equation}
where R is the Ricci scalar, $\phi$ the scalar field, $M_{Pl}$ the reduced planck mass  $M_{Pl}=\frac{m_{Pl}}{\sqrt{8\pi G}}$ and ${\mathcal L}_m$ is the Lagrangian for the matter sector, in which the matter fields $\psi_m$ are non-minimally coupled to the scalar field with
a dimensionless coupling constant  $\beta(\phi)$. 
In the chameleon and symmetron models, the properties of the single scalar field can be described by a simple, scalar field Lagrangian
\begin{equation}
\mathcal{L} = -\frac{1}{2} \left(\nabla\phi \right)^2-V(\phi)  
\end{equation}
where $V(\phi)$ is the self-interacting potential. Varying the action gives us the equations of motion for the scalar field, the Klein-Gordon equation
\begin{equation}
\label{klein}
\Box\phi = V_{eff,\phi} 
\end{equation}
where the effective potential combining the self-interaction potential and coupling term is given by
\begin{equation}
V_{eff} = V(\phi) + \frac{e^{\beta\phi/M_{pl}}\rho_m}{M_{Pl}}
\end{equation}
The chameleon screening mechanism lies in the fact that the effective mass of the scalar field calculated at the minimum, $m$, which is given by
\begin{equation}
m^2 = \frac{d^2 V_{eff}}{d\phi^2},
\end{equation}
has to be positive. For the chameleon theories, this requirement is guaranteed through the interplay between a monotonically decreasing potential $V(\phi)$ and an increasing coupling. In the symmetron model, on the other hand, the viability is restored using a ``Mexican hat" symmetry breaking potential \cite{PhysRevLett.104.231301}, the behavior of which still gives rise to a positive density-dependent mass. 

The observational consequences of such models can be demonstrated by extracting the scalar field profile, $\phi(r)$, produced by the density profile 
\be
  \rho(r)  = \left\{\def\arraystretch{1.2} 
  \begin{array}{@{}c@{\quad}l@{}}
   \rho_c  & \text{if $ r < \RR_c$ }\\
   \rho_{\infty} & \text{if $r > \RR_c$}\\
  \end{array}\right.
\ee
where $r$ is the radial distance from the center of a compact spherically symmetric configuration of density $\rho_c$ and radius $\RR_c$ (not to be confused with the Ricci scalar R), that is isolated on a uniform density background $\rho_{\infty}$. Under spherical symmetry, (\ref{klein}) becomes
\begin{equation}
\label{kleinsphere}
\frac{1}{r^2}\frac{d}{dr}\left(r^2\frac{d\phi}{dr}\right) =  \left(\frac{\partial V}{\partial \phi} + \frac{\beta(\phi) \rho_c (r)}{M_{Pl}}\right).
\end{equation}
Even though (\ref{kleinsphere}) does not have, in principle, an analytical solution, accurate approximations can be performed for two different configurations, that correspond to opposite regimes with respect to screening \cite{PhysRevLett.93.171104, PhysRevD.69.044026}. The first case is that of a large, strongly perturbing object of very large density $\rho_c$, for which the interior field is forced to acquire the value that corresponds to the minimum of the effective potential, $\phi_c$ and the scalar field profile outside the object is given by
\begin{equation}
\phi(r) = \phi_{\infty} + \frac{\left(\phi_c-\phi_{\infty}\right) \RR_c}{r}e^{-m_{\infty}r},     r > \RR_c.
\end{equation}
The corresponding fifth-force experienced by a unit mass particle outside the object is
\begin{eqnarray}
\label{force}
F_{\phi}(r) & =& 2\beta_{\infty}^2\left(\frac{\Delta \RR_c}{\RR_c}\right)\frac{GM}{r^2}\left(1+m_{\infty}r\right)e^{-m_{\infty}r},            
\end{eqnarray}
where $m_{\infty},\beta_{\infty}$ are respectively the background values of the mass and coupling and M the mass of the object. Given the characteristic large values of the Compton wavelength $\lambda_c \equiv m_{\infty}^{-1}$, the scalar field is essentially free within our scales of interest and the Yukawa suppression can be neglected in (\ref{force}),
\begin{eqnarray}
\label{screenapprox}
F_{\phi}(r)  & \approx & 2\beta_{\infty}^2\left(\frac{\Delta \RR_c}{\RR_c}\right)\frac{GM}{r^2},  \hspace{1cm}m_{\infty}r \ll 1.
\end{eqnarray}
The above approximation is valid when the ``screening factor" is
\begin{equation}
\label{fac}
\frac{\Delta \RR_c}{\RR_c} = \frac{|\phi_{\infty}-\phi_{c}|}{2\beta_{\infty}M_{Pl} \Phi_N} \ll 1,
\end{equation}
which also defines the criterion for the existence of a thin shell \cite{PhysRevLett.104.231301,Khoury:2013tda}, whose mass is the fraction of the total that actually contributes to the fifth force, due to the strong Yukawa suppression deep inside dense objects. The Newtonian gravitational potential is denoted by $\Phi_N$ in (\ref{fac}). On the other hand, when linear perturbation theory is valid, which is the case when $\frac{\Delta \RR_c}{\RR_c} > 1$, the linearized form of (\ref{kleinsphere}) gives 
\begin{eqnarray}
\label{forcelin}
F_{\phi}(r) & \approx & 2\beta_{\infty}^2\frac{GM}{r^2}, \hspace{1cm}m_{\infty}r \ll 1            
\end{eqnarray} 
for the fifth force. Based on (\ref{screenapprox})-(\ref{forcelin}), we see that in the linear regime the fifth force is the same as the Newtonian force with a coupling $2 \beta_{\infty}^2$ and deep in the non-linear (screened) regime, it is suppressed by the thin shell factor (\ref{fac}).

Furthermore, it should be also noted that, as shown in \cite{Brax:2012gr}, one can derive a pair of functions $\beta(a),m(a)$, for the characterization of a model within the above framework. Unlike models with constant couplings, symmetrons exhibit an additional form of screening \cite{PhysRevLett.104.231301,Olive:2007aj} due to the fact that in dense environments symmetry is restored and the coupling $\beta(\phi)$ vanishes.

Adopting this formulation, linear perturbation theory gives \cite{PhysRevD.80.044027,Brax:2012gr} for the growth of CDM density perturbations in the quasi-static limit and for sub-horizon scales 
\begin{equation}
\label{growth}
\ddot{\delta}_m + 2H\dot{\delta}_m = \frac{3}{2} \Omega_m(a)H^2 \delta_m\frac{G_{eff}(k,a)}{G}
\end{equation}
with 
\begin{equation}
\label{effective}
\frac{G_{eff}(k,a)}{G} = 1+ \frac{2\beta^2(a)k^2}{k^2 + a^2 m^2(a)}
\end{equation}
where $a$ is scale factor, with $a=1$ today, and $k$ is the comoving wavenumber.

The effects of gravity modifications at the linear approximation are incorporated in the second term. For very large scales and/or early times (GR regime), $am(a)/k \gg1$ and (\ref{growth}) reduces to the standard GR  expression in the weak gravity regime, where the Newtonian gravitational potential is given by the Poisson equation,
\begin{equation}
\label{poisson}
\nabla^2\Phi_N =  \frac{3}{2} \Omega_{m0}\frac{H_0^2}{a}\delta_m.
\end{equation}
When $am(a)/k \le1$ however (scalar-tensor regime), the second term becomes significant and gives the linearized Klein-Gordon equation for the fifth potential $\phi$
\begin{equation}
\label{fourklein}
\phi(k,a)=-\frac{\beta(a)}{k^2+a^2m^2(a)}\frac{\bar{\rho}_m a^2}{M_{pl}} \delta_m.
\end{equation}
with the real space expression being
\begin{equation}
\label{realklein}
\nabla^2\phi = a^2m^2(a)\phi + \frac{\beta(a) a^2 \bar{\rho}_m}{M_{pl}}\delta_m.
\end{equation}

\subsubsection{The $f(R)$ model}
$f(R)$ theories \cite{Carroll:2003wy} are widely-studied modified gravity scenarios,  that give rise to acceleration on cosmic scales and can be incorporated \cite{Brax:2008hh} into the chameleon formalism with a constant coupling $\beta=1/\sqrt{6}$ . The first model we tested thus, was the Hu-Sawicky $f(R)$ model \cite{Hu:2007nk} with a scalar field mass
\begin{equation}
m(a) = \left(\frac{1}{3(n+1)}\frac{\bar{R}}{|\bar{f}_{R_0}|}\left(\frac{\bar{R}}{\bar{R}_0}\right)^{n+1}\right)^{\frac{1}{2}}
\end{equation}
where
\begin{equation}
\bar{R}= -3(H_0^2\Omega_{m0})^2\left(a^{-3}+4\frac{\Omega_{\Lambda0}}{\Omega_{m0}}\right)
\end{equation}
where $H_0$ is the Hubble Constant and $\Omega_{\Lambda0}$ and $\Omega_{m0}$ are, respectively, the dark energy and dark matter fractional energy densities today. The mass takes the form
\begin{equation}
m(a)=\frac{1}{2997}\left(\frac{1}{2|\bar{f}_{R_0}|}\right)^{\frac{1}{2}}\frac{\left(\Omega_{m0} a^{-3}+4\Omega_{\Lambda0}\right)^{1+\frac{n}{2}}}{\left(\Omega_{m0}+4\Omega_{\Lambda0}\right)^{\frac{n+1}{2}}} [Mpc/h].
\end{equation}
Furthermore, the screening factor is given by
\begin{equation}
\frac{\Delta \RR_c}{\RR_c} = \frac{3}{2} \abs{\frac{\bar{f}_{R_0}}{\Phi_N}}\left(\frac{\Omega_{m0}+4\Omega_{\Lambda0}}{\Omega_{m0} a^{-3}+4\Omega_{\Lambda0}}\right)^{n+1}.
\end{equation}
$\bar{f}_{R_0}=\frac{df(R)}{dR}\big|_{z=0}$ and $n$ are the model's free parameters. In this paper, we
consider the model for $n=1$ and $\abs{\bar{f}_{R_0}}=\{10^{-4},10^{-5},10^{-6}\}$. These describe cosmologically viable  scenarios whose non-linear properties have been simulated using the full Klein-Gordon equation  \cite{Zhao:2010qy,Winther:2014cia} with which our results can be compared.
\subsubsection{The symmetron model}
\label{symsec}

The general framework laid previously, can also incorporate the symmetron model, with a ``Mexican hat" symmetry breaking potential \cite{PhysRevLett.104.231301}, for which scalar fields couple to matter after $a>a_{ssb}$, with
\begin{equation}
\label{symscreen}
\begin{split}
m(a) & = \frac{1}{\lambda_{\phi0}}\sqrt{1-\left(\frac{a_{ssb}}{a}\right)^3} \\
\beta(a) & = \beta_0\sqrt{1-\left(\frac{a_{ssb}}{a}\right)^3}
\end{split}
\end{equation}
and the coupling vanishes for $a<a_{ssb}$, when symmetry is restored. The screening factor for this model becomes \citep{Davis:2011pj,Winther:2014cia}
\begin{equation}
\frac{\Delta \RR_c}{\RR_c} = \frac{\Omega_{m0}}{3.0 a_{ssb}^3}\left(\frac{\lambda_{\phi 0}}{Mpc/h}\right)^2\abs{\frac{10^{-6}}{\Phi_N}}
\end{equation}
We consider this model with values $a_{ssb}=0.5,\beta_0=1$ and $\lambda_{\phi 0}=1Mpc/h$ which again have been shown \cite{Davis:2011pj} to predict deviations consistent with experimental constraints. It should be also pointed out that, as explained previously, models of this type exhibit field dependent couplings which cause additional screening due to the coupling suppression in dense environments, where symmetry is again restored. This effect is not taken into account in our approximate scheme. 
\subsection{Simulating non-linear clustering}
\label{sec:Non-linear approaches}

\subsubsection{The N-body method}
\label{sec:Nbody}
\begin{figure*}[!t]
\bc
{\includegraphics[width=0.48\textwidth]{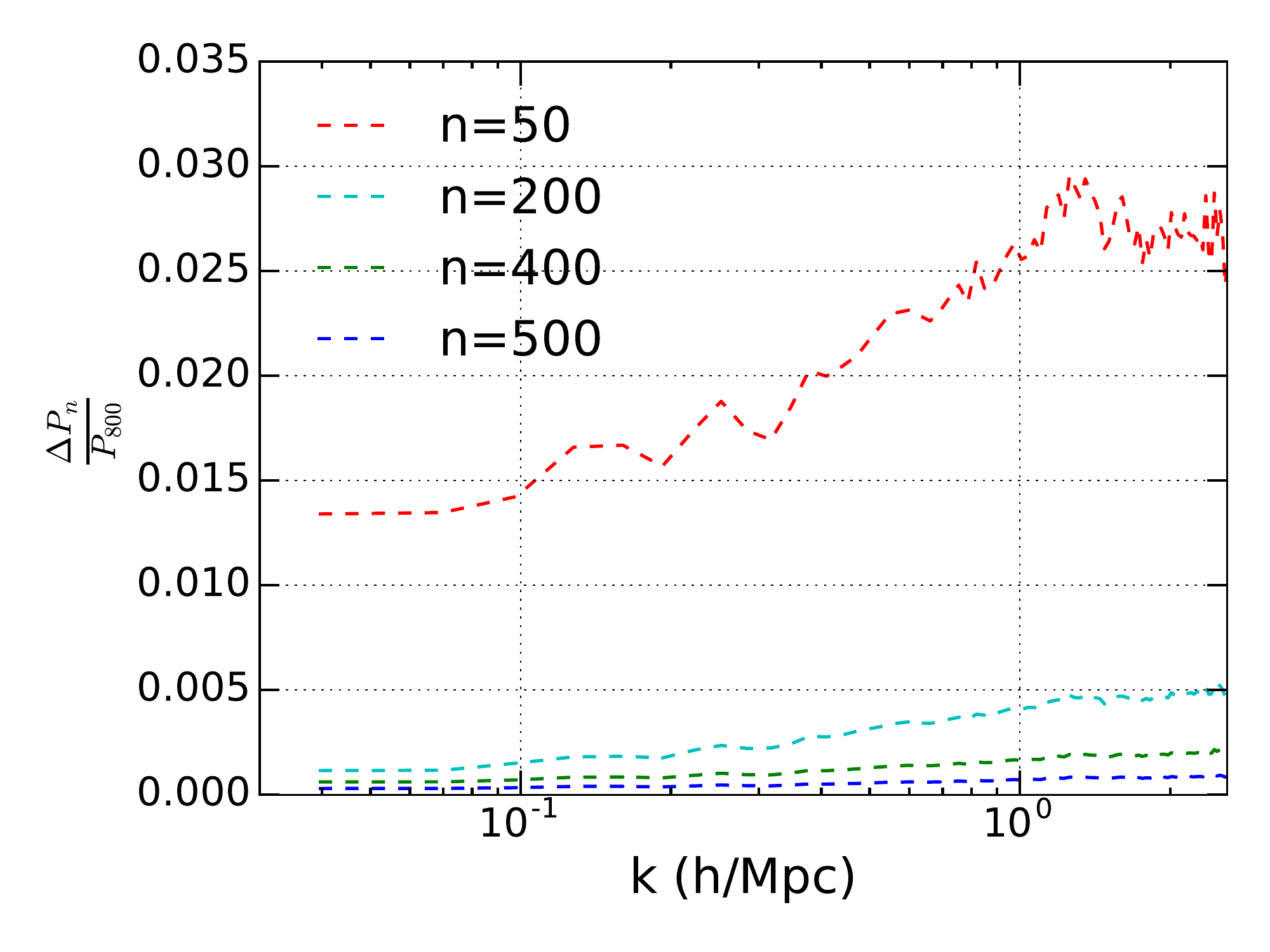}
\includegraphics[width=0.48\textwidth]{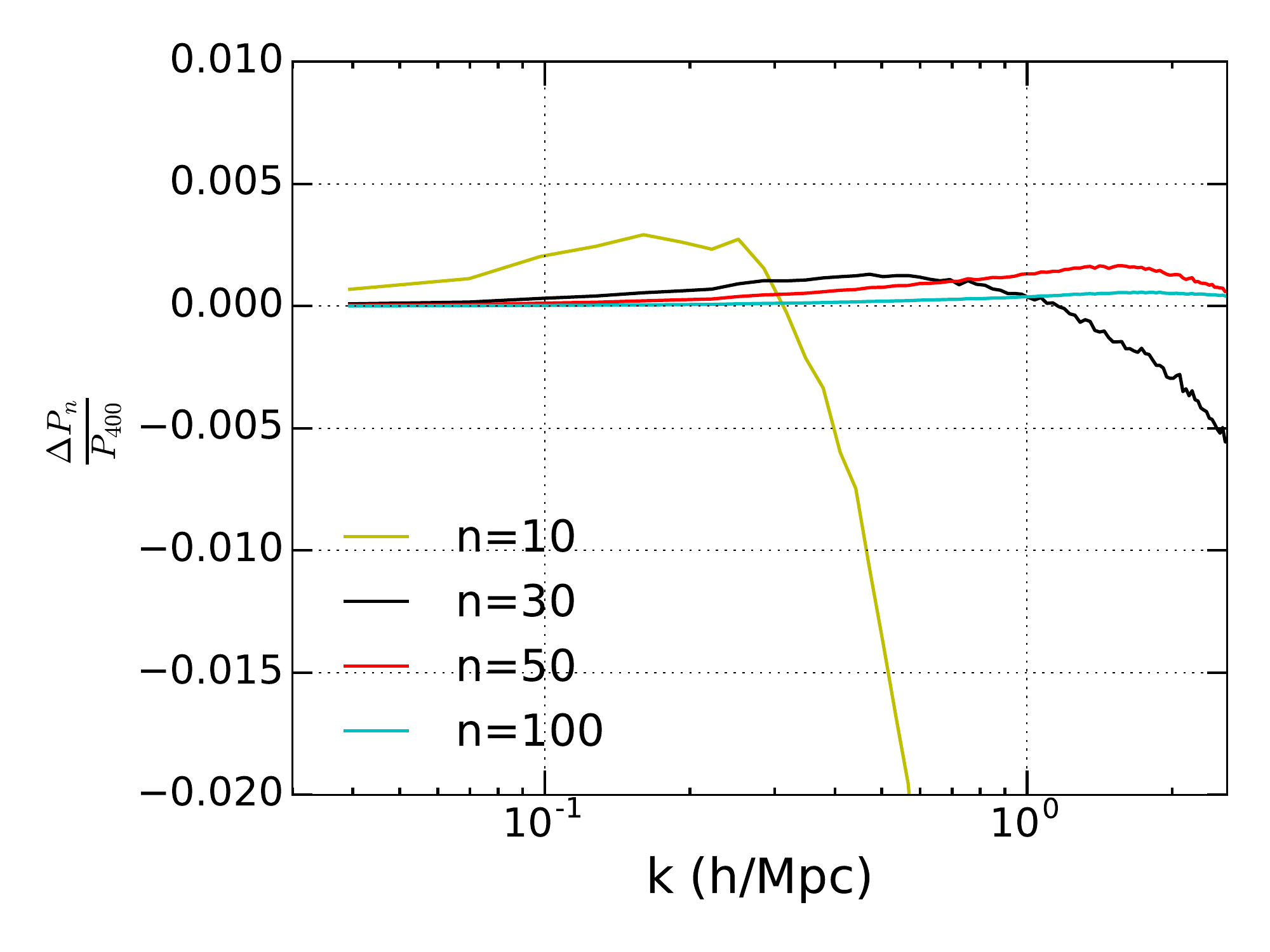}
}
\caption{The fractional difference between the $\Lambda$CDM power spectrum, $P_{n}$, for one realization as obtained using different choices of time steps, $n$, and the high resolution results, $P_{800}$ for PM (left) , and $P_{400}$ for COLA (right), respectively.}
\label{figp}
\ec
\end{figure*}

The COLA code has been loosely based on A. Klypin's PM code \cite{Klypin:1997sk}, and this motivates the latter's use as a comparison for our approximate scheme's effectiveness. 
It is also a simple and representative implementation of a Particle-Mesh (PM) N-body code. N-body simulations for MG using the PM code have been performed previously \cite{Stabenau:2006td,Laszlo:2007td,Khoury:2009tk}. For each scenario, we consider  10  simulated realizations, initialized at an initial redshift $z_i=49$, at which density perturbations on the scales we study are linear. After providing a linear power spectrum from the cosmological code CAMB \cite{Lewis:1999bs} for the desired $\Lambda$CDM cosmology at the time $z_i$, $N_p=256^3$ particles are placed in our simulation box with side L=200 Mpc/h, in a mesh of $512^3$, using 1st order Lagrangian Perturbation Theory (Zel'dovich approximation) \cite{Zeldovich:1969sb}. The parameters that define our background $\Lambda$CDM cosmology are $\Omega_{m0}=0.25$, $\Omega_{\Lambda0}=0.75$, $h=0.7$, $n_s=1.0$ and $\sigma_8=0.8$. The particle positions are updated, using 500 time steps, through the displacement equation:
\begin{equation}
\label{dispL}
\ddot{\mathbf{x}} + 2H \dot{\mathbf{x}} = -\frac{1}{a^2}\nabla_{\mathbf{x}} \Phi_N.
\end{equation}
In Fig. \ref{figp}, it is shown that the choice of 500 iterations, which corresponds to steps of $\Delta a=0.00196$ in the scale factor, guarantees convergence at the 0.08\% level.

In MG cosmologies, the modified geodesic equation gives, in the weak gravity regime, the modified version of (\ref{dispL}),  
\begin{equation}
\label{dispMG}
\ddot{\mathbf{x}} + \left(2H+\frac{\beta}{M_{Pl}}\dot{\phi}\right) \dot{\mathbf{x}} = -\frac{1}{a^2}\left(\nabla_{\mathbf{x}} \Phi_N + \frac{\beta}{M_{Pl}}\nabla_{\mathbf{x}} \phi \right),
\end{equation}
where the term $\abs{\frac{\beta}{M_{Pl}}\dot{\phi}}$ is negligible given observational constraints from variations of constants \cite{Winther:2014cia}. Equation (\ref{dispMG}), which also holds for the full non-linear KG description, forms a closed system of equations with  (\ref{poisson}) and (\ref{realklein}) that are solved in the Fourier space for the potentials $\Phi_N$ and $\phi$. 

The linearized form of KG equation, (\ref{realklein}), does not incorporate the screening effects. To account for the screening effect, we adopt an effective parameterization similar to the one proposed in \cite{Winther:2014cia}. In section Sec. \ref{sec:Modified gravity and screening models}, we showed that the linear solution for the fifth force, (\ref{screenapprox}), is suppressed by the screening factor deep in the non-linear regime. As a result, we incorporate the screening effects by explicitly attaching the screening factor to the fifth force in accordance with (\ref{screenapprox})-(\ref{forcelin}) and (\ref{dispMG}), 
\begin{equation}
\label{dispscreen}
\ddot{\mathbf{x}} + 2H \dot{\mathbf{x}}=-\frac{1}{a^2}\left(\nabla_{\mathbf{x}} \Phi_N+ \frac{\Delta \RR_c}{\RR_c}  \frac{\beta}{M_{Pl}}\nabla_{\mathbf{x}} \phi \right).
\end{equation}
To interpolate properly between the screened and the unscreened regime we set
\begin{equation}
\label{sfac}
  \frac{\Delta \RR_c}{\RR_c}  = \left\{\def\arraystretch{1.2} 
  \begin{array}{@{}c@{\quad}l@{}}
   \frac{\phi(a)}{2\beta(a) M_{Pl} \abs{\Phi_N}}  & \text{if $ \frac{\phi(a)}{2\beta(a) M_{Pl} \abs{\Phi_N}} <1$ }\\
    1 & \text{if $\frac{\phi(a)}{2\beta(a) M_{Pl} \abs{\Phi_N}}>1.$}\\
  \end{array}\right.
\end{equation}
Within our approximate scheme, the functions $|\phi_{\infty}-\phi_{c}|$ and $\beta_{\infty}$ have been set equal to the background ones $|\phi(a)|$ and $\beta(a)$ correspondingly, which has been shown to be a good approximation in \citep{Winther:2014cia}.
\subsubsection{The COLA method}
\label{sec:COLA}

\begin{figure*}[!t]
\bc
{\includegraphics[width=0.48\textwidth]{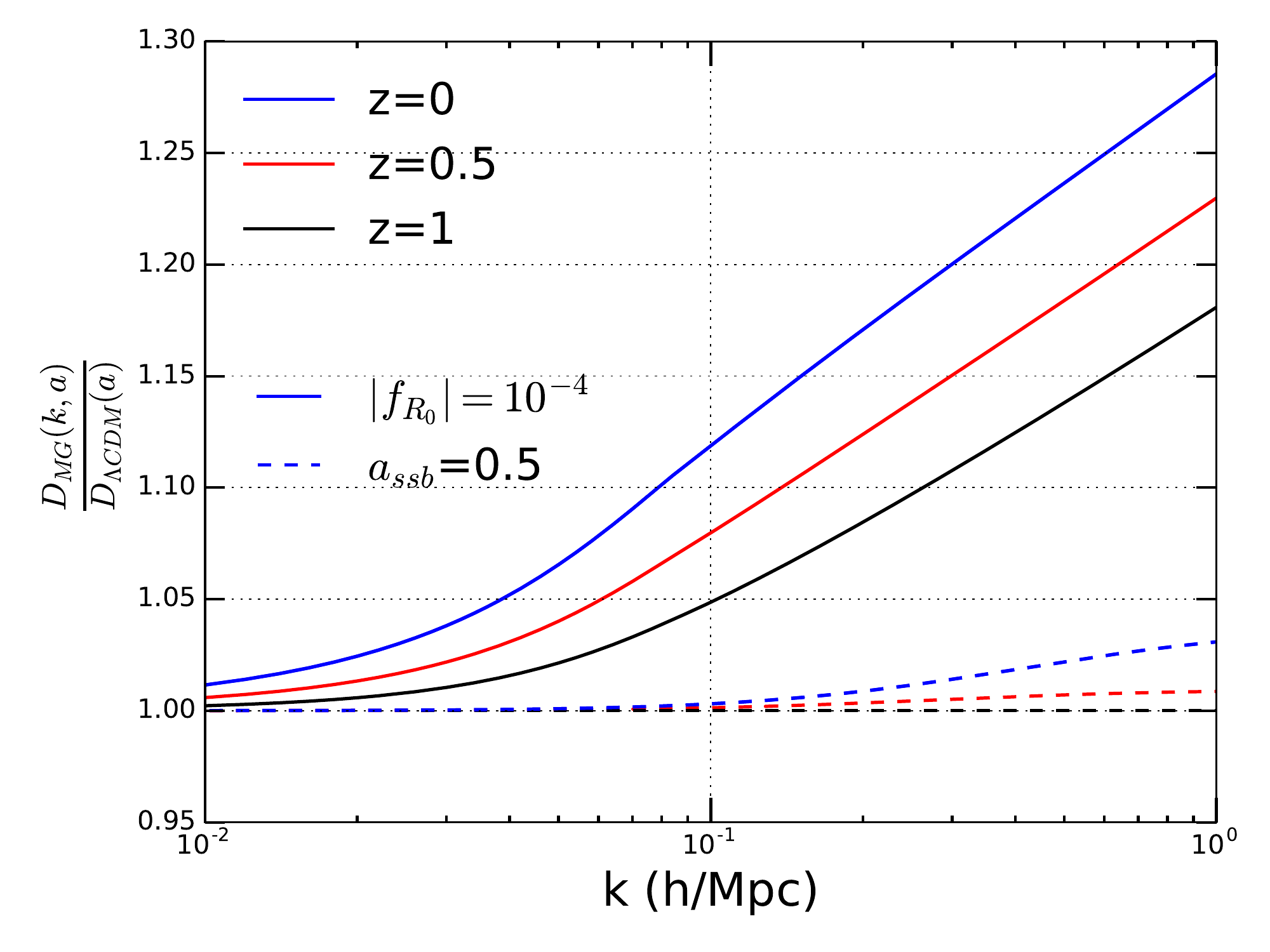}
\includegraphics[width=0.48\textwidth]{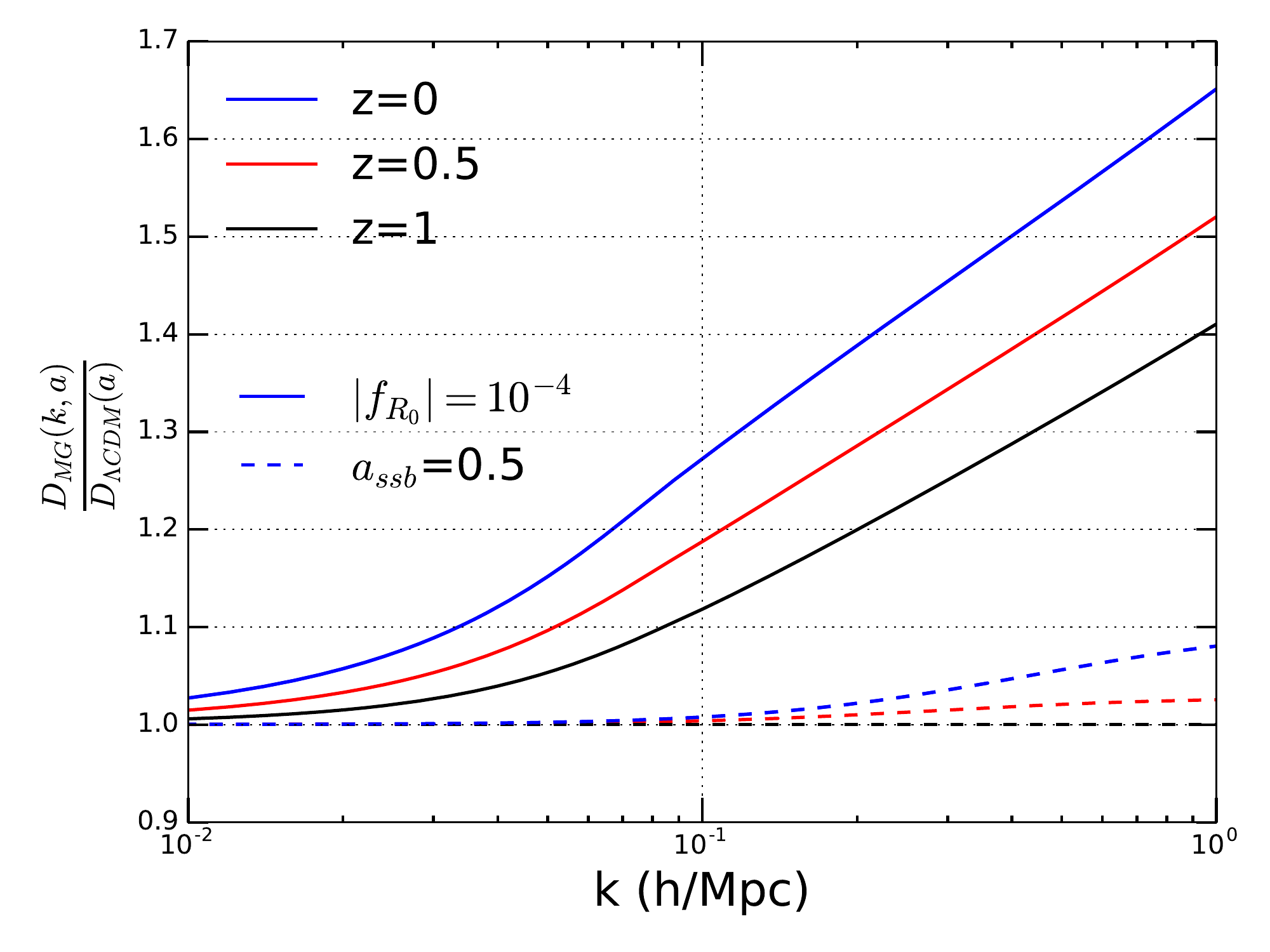}
}
\caption{The ratio of the $1^{st}$ (left) and $2^{nd}$ (right) order growth factors, $D_1$ and $D_2$ respective, relative to the $\Lambda$CDM growth factor in each case, for the $\abs{f_{R_0}}=10^{-4}$ model (solid lines) and the symmetron scenario (dashed lines) described in section \ref{symsec}.}
\label{fig1}
\ec
\end{figure*}
The fact that N-body codes manage to simulate the Large Scale Structure accurately but at a significant computational cost, has motivated the development of several analytical perturbative techniques to avoid a full blown N-body simulation. Lagrangian Perturbation Theory (LPT) \cite{Zeldovich:1969sb,Bouchet:1994xp} works perturbatively in a Lagrangian displacement field and manages to give accurate results in the Linear and the Mildly Non-Linear regime. However, it quickly fails to capture the non-linearities associated with the smaller scales and consequently it underestimates significantly the power at large $k$. Given that we have to choose between accurate, but expensive N-body simulations and fast but approximate perturbative techniques, it is reasonable to ask whether one can efficiently combine the benefits of both approaches. Such a hybrid method, named COmoving Lagrangian Acceleration (COLA) was proposed in \cite{Tassev:2013pn}. Here we outline the basic framework and its modifications for MG, while details can be found at \cite{Tassev:2013pn}. The particle comoving positions are decomposed as a sum of two pieces, in the ``manifestly" exact form 
\begin{equation}
\label{basx}
\mathbf{x} = \mathbf{x}_{res} + \mathbf{x}_{LPT}
\end{equation}
By defining a new time variable $d\theta \equiv H_0\frac{dt}{a^2}=\frac{H_0}{a}d\eta$, where $\eta$ is conformal time, (\ref{dispL}) can be cast in the simpler form 
\begin{equation}
T^2(\mathbf{x})=-\frac{a^2}{H_0^2}\nabla_{\mathbf{x}} \Phi_N,
\end{equation}
with $T\equiv \frac{d}{d\theta}=\frac{a}{H_0}\partial_{\eta}=Q(a)\partial_a$, and $Q(a)=a^3\frac{H(a)}{H_0}$. In the Lagrangian description $\mathbf{x}=\mathbf{q}+\mathbf{s}(\mathbf{q},a)$, with $\mathbf{q}$ the initial Eulerian position and $\mathbf{s}$ the Lagrangian displacement and
\begin{equation}
T^2(\mathbf{s})=-\frac{a^2}{H_0^2}\nabla_{\mathbf{x}} \Phi_N.
\end{equation}
One can now solve for the residual displacement in $\Lambda$CDM
\begin{equation}
T^2(\mathbf{s}_{res})=-\frac{a^2}{H_0^2}\nabla_{\mathbf{x}} \Phi_N - T^2[D_1(a)] \mathbf{s}_1 - T^2[D_2(a)] \mathbf{s}_2,
\end{equation}
where $D_1(a)$ and $D_2(a)$ are the first and second order growth factors, respectively,  and $\mathbf{s}_1$, $\mathbf{s}_2$ are the Zel'dovich and second order LPT displacements. 
The fact that the LPT piece is evolved analytically and we only solve numerically for $\mathbf{s}_{res}$, can be interpreted as working on a frame that is co-moving with observers that follow LPT trajectories.

\begin{figure*}[!tb]
\bc
{\includegraphics[width=0.48\textwidth]{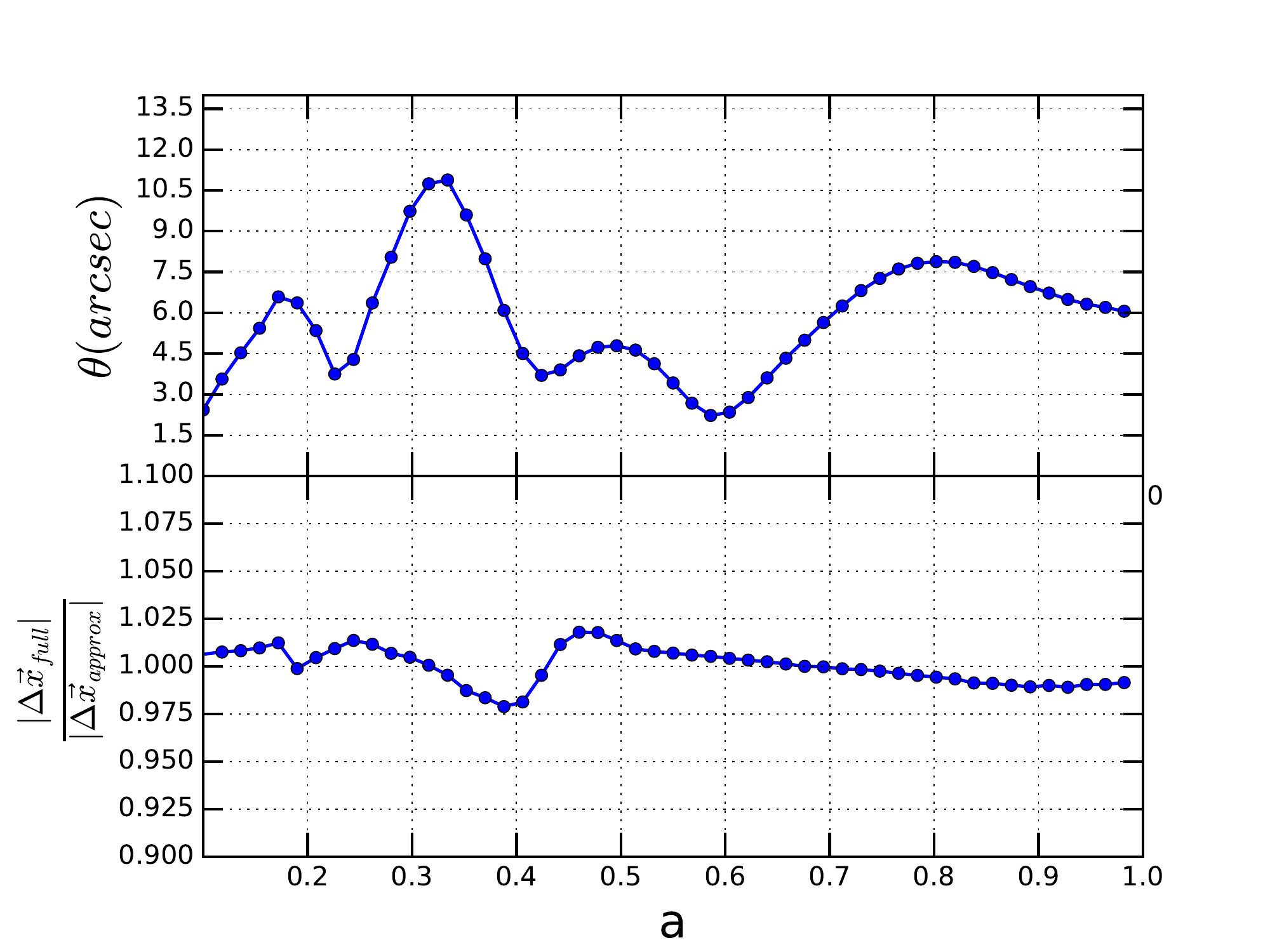}
 \includegraphics[width=0.48\textwidth]{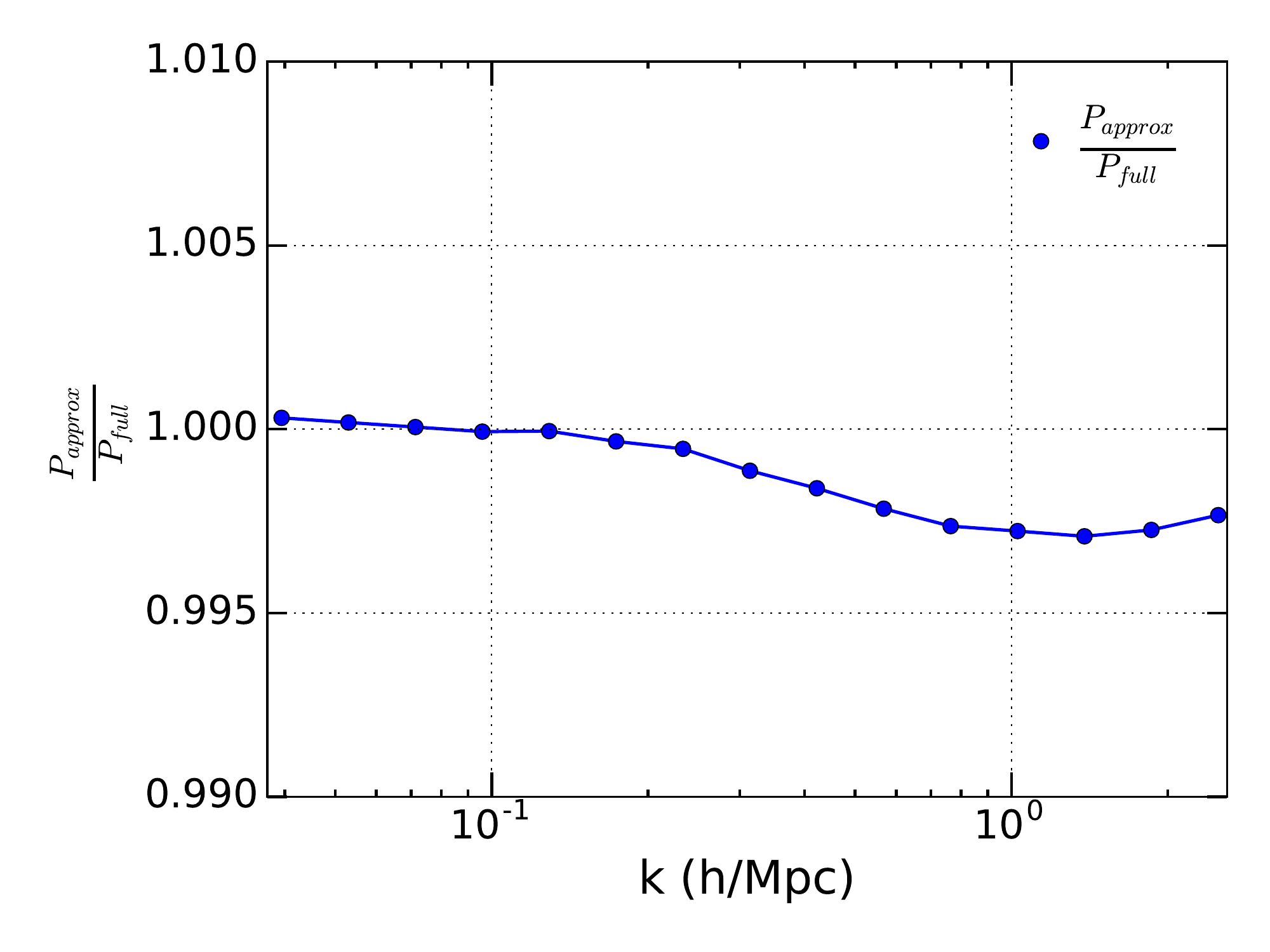}
}
\caption{Left: Tracking a particle's position throughout the simulation for the exact and approximate hybrid in the $\abs{f_{R_0}}=10^{-4}$ model. Right: Ratio of the power spectra obtained from the approximate and the exact method for the $\abs{f_{R_0}}=10^{-4}$ model today. }
\label{fig2}
\ec
\end{figure*}

$T$ can be discretized using a Leapfrog scheme \cite{Quinn:1997iy} to get the core COLA equations for each particle's position and velocity change between the times $a_i,a_f$
\begin{equation}
\label{fullMG}
\begin{split}
\mathbf{x}(a_f) & = \mathbf{x}(a_i) + \mathbf{\upsilon}(a_c)\int_{a_i}^{a_f}\frac{da}{Q(a)} + \\
                   & + \left(\mathbf{s}_1(\mathbf{q},a_f)-\mathbf{s}_1(\mathbf{q},a_i)\right)+\\
                   & + \left(\mathbf{s}_2(\mathbf{q},a_f)-\mathbf{s}_2(\mathbf{q},a_i)\right) \\
\mathbf{\upsilon}(a_f) & = \mathbf{\upsilon}(a_i) - \left(\int_{a_i}^{a_f}\frac{a}{a_cQ(a)}da\right) \times \\
                              & \left[ - 1.5\Omega_{m0} a_c\left(\nabla_{\mathbf{x}} \tilde{\Phi}_{N}(\mathbf{x})+\frac{\Delta \RR_c}{\RR_c}\frac{\beta}{M_{Pl}}\nabla_{\mathbf{x}} \tilde{\phi}(\mathbf{x})\right) - \right. \\
                              & \left. - T^2[\mathbf{s}_1](a_c)-T^2[\mathbf{s}_2](a_c)\right],
\end{split}                              
\end{equation}
 where  a tilde denotes a quantity in units of $1.5 \Omega_{m0}H_0^2/a$.
 
Initial conditions are produced using the 2LPT initial conditions code (2LPTic) \cite{Scoccimarro:1997gr} which does so by performing LPT up to second order. In a $\Lambda$CDM cosmology, growth functions $D_1(a)$ and $D_2(a)$ are scale independent \cite{Zeldovich:1969sb,Bouchet:1994xp} 
 and one only needs to produce an LPT snapshot for $z=0$ for both generating initial conditions and obtaining the LPT terms at the different timesteps. In such a case, the LPT displacements are given by $\mathbf{s}_{1}(\mathbf{q},a)=D_{1}(a)\mathbf{s}_{1}(\mathbf{q},a_0)$, $\mathbf{s}_{2}(\mathbf{q},a)=D_{2}(a)\mathbf{s}_{2}(\mathbf{q},a_0)$ and (\ref{fullMG}) reduces to the standard COLA $\Lambda$CDM scheme (with the fifth force term omitted). Initial conditions and background cosmology are produced, for 10 realizations, for the same cosmological parameters as used in the PM code, at the  initial redshift z=9.0 which has been shown \cite{Tassev:2013pn} to work well for COLA in $\Lambda$CDM. The simulation box size, number of particles and mesh size are  the same as used in the PM code. It should be noted though that we don't perform a comparison of the codes by initiating both with identically seeded initial conditions, but instead, we  compare the statistical consistency of the means of the 10 runs for each of the two techniques with the respective sets each using different random generated seeds. In its initial formulation, COLA was used with 10 time steps, which enables accurate predictions down to $k\sim 0.5 $ h/Mpc, which can be also seen in Fig. \ref{figp}, where the $\Lambda$CDM power spectrum by COLA is presented for various choices of time steps. By increasing the number of steps to 50, still significantly fewer than the typical number of iterations performed in a standard N-body code, we can provide accuracy down to smaller scales, $k \sim2 $ h/Mpc. The $\Lambda$CDM COLA run-time in this set up is $\sim$10 times shorter than that of the PM code. COLA's accuracy as a function of the number of time steps used, is further discussed in section \ref{sec:Modified gravity results}.

When gravity is modified, the core equations need to be changed appropriately to account for the additional fifth forces and the screening effects. As in the N-body code, one can use an approximate framework to model the modified gravity effects both on the growth rate and screening:  solving the linearized KG equation (\ref{fourklein}) and attaching the screening factor (\ref{sfac}) to fifth force term in (\ref{fullMG}). For the LPT component of COLA, one must consider that, in MG theories, the growth factors $D_1$ and $D_2$ become scale dependent. 

In Figure \ref{fig1}, we summarize the first and second order growth factors for a chameleon and symmetron scenario. In each case as we approach late times, $z<2$ for the chameleon model, and $z<0.5$ for the symmetron, we find significant scale dependent deviations from $\Lambda$CDM at the level of 10\% for $D_1$ and 25\% for $D_2$ at k=0.1 h/Mpc today, which means that not all Fourier modes evolve the same way with time \cite{Valkenburg:2015dsa}. This causes the LPT trajectories of a given particle to bend, in principle. As a consequence one has to be very cautious about how to obtain the LPT terms at the different times. We briefly outline the application of LPT to scale dependent growth functions in MG in appendix \ref{App:AppendixA}. 

Unlike the $\Lambda$CDM case, here the growth factors' scale dependent nature does not allow one to evolve the Zel'dovich and $2^{nd}$ order displacements with a single scale independent function for all scales. To account for that, we have considered two alternative modifications to COLA. In the first approach, we  create an MG version of COLA that 
calculates the LPT displacements numerically at each time step using an MG version of 2LPTic. The relevant LPT terms in (\ref{fullMG}) are calculated after Fourier transforming (\ref{fourdisps}) and (\ref{fouraccels}). Besides the modified N-body component, the fact that we have to solve numerically for the Lagrangian terms at every discrete time step increases the computational cost significantly. In a second approach, we utilize  the fact that the LPT part of the scheme serves to evolve the linear scales, for which the MG deviations with respect to $\Lambda$CDM are known to be small for most times and adopt an approximate scheme in which only the N-body part is modified and the $\Lambda$CDM solutions are used for the Lagrangian displacements. The resulting scheme has the same N-body component as in (\ref{fullMG}) and the known $\Lambda$CDM LPT terms, in which the Lagrangian displacements are evolved with $D_{1,\Lambda}(a)$ and $D_{2,\Lambda}(a)$.
\begin{equation}
\label{approxMG}
\begin{split}
\mathbf{x}(a_f) & = \mathbf{x}(a_i) + \mathbf{\upsilon}(a_c)\int_{a_i}^{a_f}\frac{da}{Q(a)} + \\
                   & + \left(D_{1,\Lambda}(a_f)-D_{1,\Lambda}(a_i)\right)\mathbf{s}_{1}(\mathbf{q},a_0)+ \\
                   & + \left(D_{2,\Lambda}(a_f)-D_{2,\Lambda}(a_i)\right)\mathbf{s}_{2}(\mathbf{q},a_0) \\
\mathbf{\upsilon}(a_f) & = \mathbf{\upsilon}(a_i) - \left(\int_{a_i}^{a_f}\frac{a}{a_cQ(a)}da\right) \times \\
                              & \left[ - 1.5\Omega_{m0} a_c\left(\nabla_{\mathbf{x}} \tilde{\Phi}_{N}(\mathbf{x})+\frac{\Delta \RR_c}{\RR_c}\frac{\beta}{M_{Pl}}\nabla_{\mathbf{x}} \tilde{\phi}(\mathbf{x})\right) - \right. \\
                              & \left. - T^2[D_{1,\Lambda}](a_c)\mathbf{s}_{1}(\mathbf{q},a_0)-T^2[D_{2,\Lambda}](a_c)\mathbf{s}_{2}(\mathbf{q},a_0)\right].                                        
\end{split}
\end{equation}

A comparison of the two approaches for the $f(R)$ model with $\abs{f_{R_0}}=10^{-4}$, for which we expect the largest modifications, is shown in Fig. \ref{fig2}.  One comparison  tracks a given particle inside our volume during the simulation and we also compare the resulting power spectra  using both schemes. We find excellent agreement between the approximate and fully modified schemes, in both the linear and mildly non-linear regimes. We also find very small differences for the position and displacement vectors (magnitude \& direction) with differences in angular orientation of at most $11$ arcseconds, and differences in the magnitude of steps less than $2,5\%$, which result in power spectra that have a fractional difference no larger  than 0.3\% today. The approximate scheme takes under half the run time of the full implementation. In light of these results, we adopt the approximate scheme in the COLA simulations used in this analysis. This has the great advantage of not having to solve numerically for the LPT displacements at every time step, without sacrificing much accuracy.  

\section{Analysis/Results}
\label{sec:Analysis/Results}

\begin{figure}[!tb]
\bc
{\includegraphics[width=0.48\textwidth]{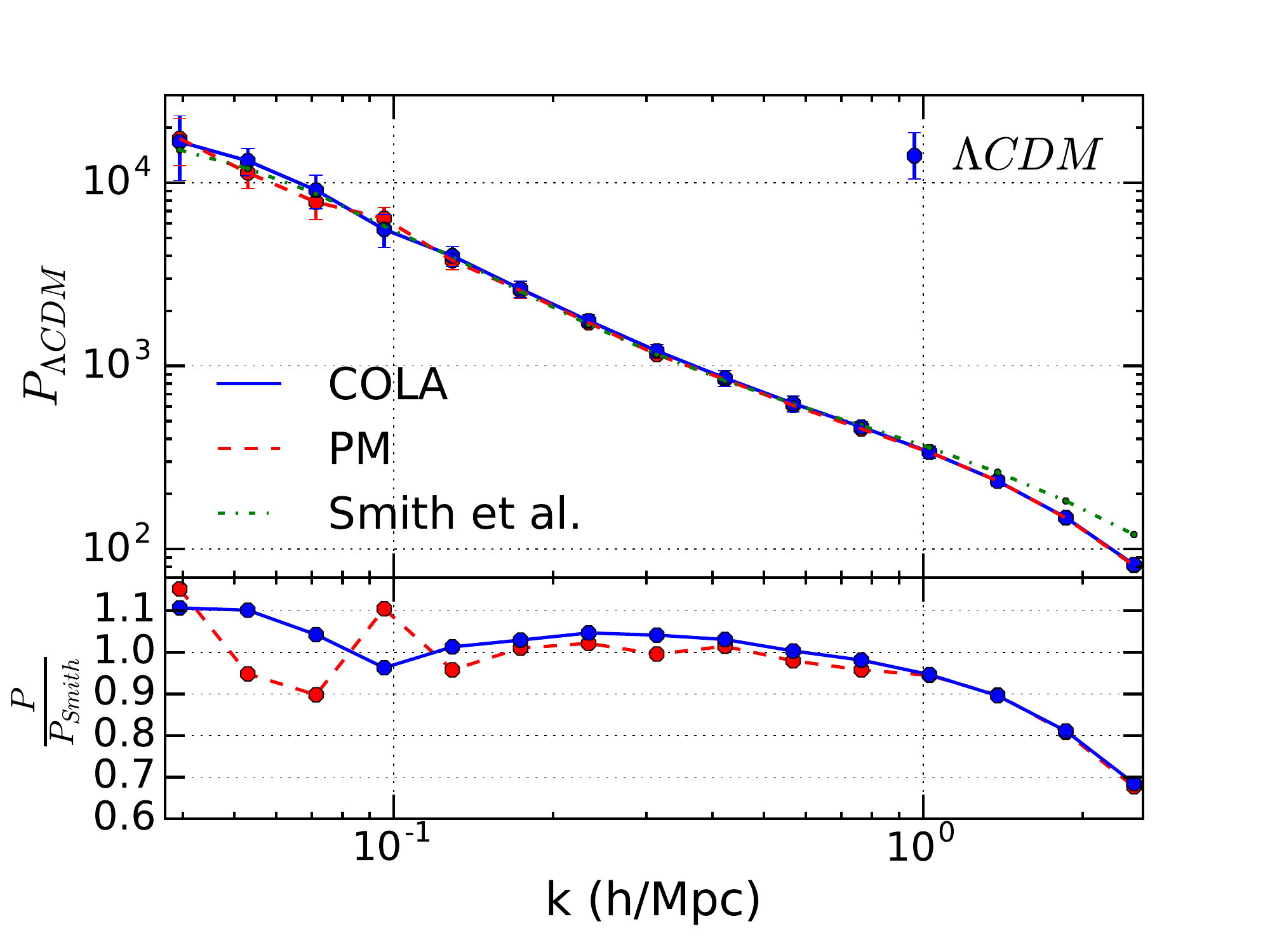}
}
\caption{[Top] Power spectra benchmarking for $\Lambda$CDM with the PM N-body code [red dashed line] and COLA method [blue full line]. The nonlinear power spectrum fit developed by Smith et al. [green dotted line] is also shown for comparison. [Bottom] Ratio between both the COLA and PM code $\Lambda$CDM results above to the fit by Smith et al. The number of time steps used for COLA and PM is 50 and 500, respectively.}
\label{fig3}
\ec
\end{figure}
\begin{figure*}[!tb]
\bc
{
\includegraphics[width=0.48\textwidth]{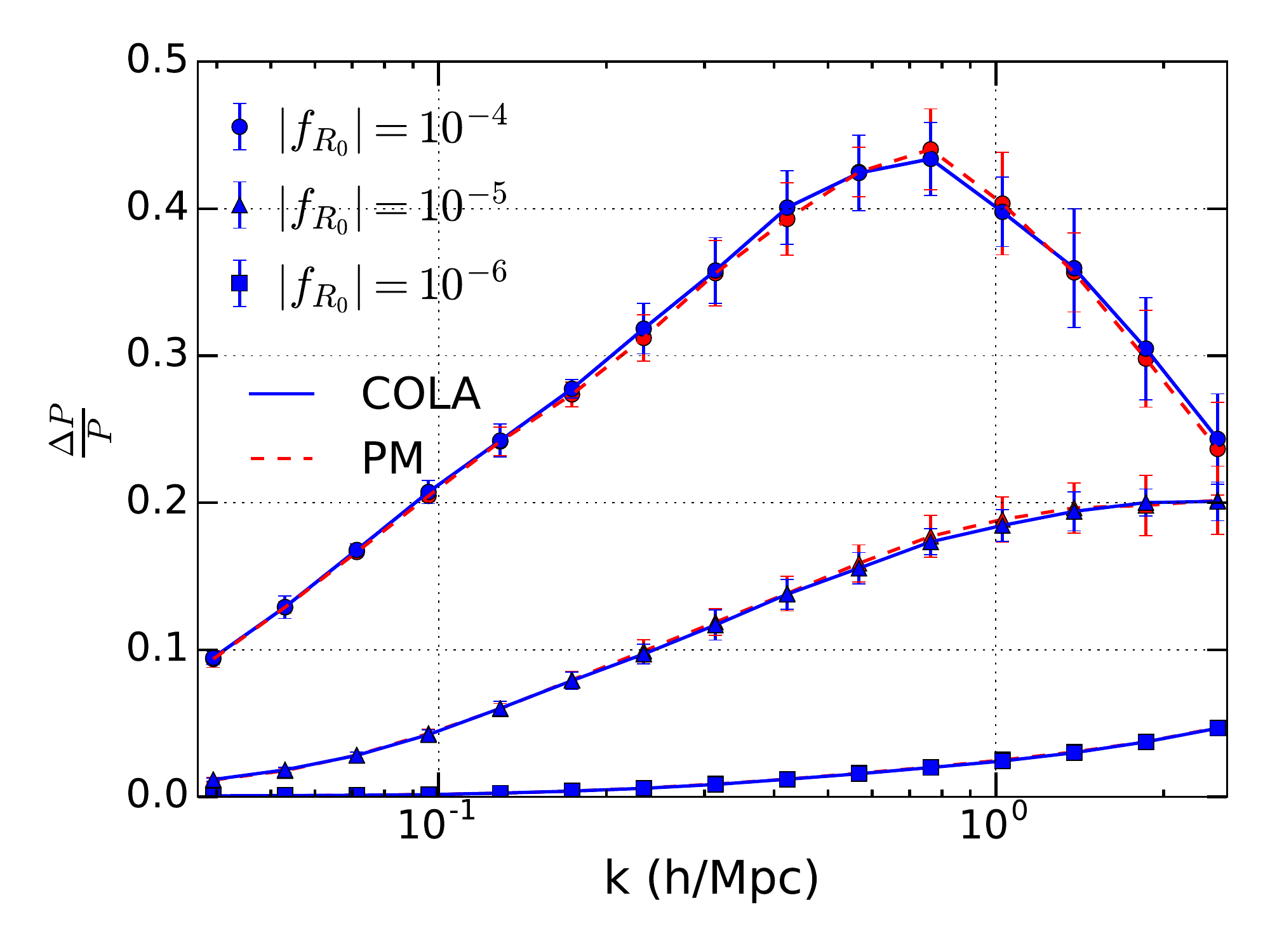}
\includegraphics[width=0.48\textwidth]{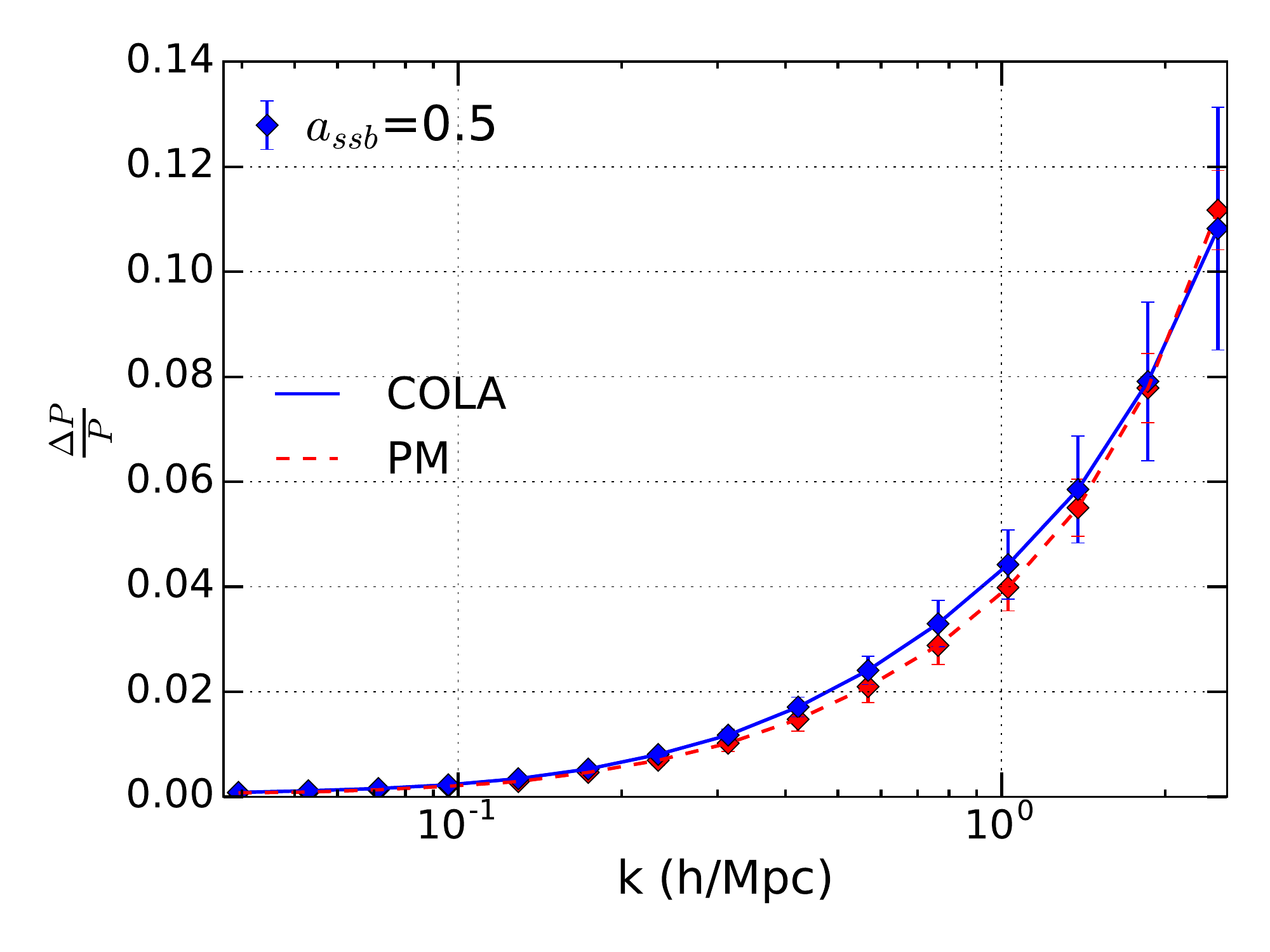}
}
\caption{Fractional difference in the CDM power spectra for the MG scenario relative to the $\Lambda$CDM model, $\frac{\Delta P}{P_{\Lambda CDM}}$, at $a=1$, for the same initial conditions for [left] the $f(R)$ scenario, for $f_{R_0}=10^{-4},10^{-5}$ and $10^{-6}$, and [right]  the symmetron model with $a_{ssb}=0.5$. The averaged results, and standard deviations, from the simulations with the PM code [red dashed line] and the COLA code [blue full] are presented in each case. The number of time steps used for COLA and PM is 50 and 500, respectively.}
\label{fig4}
\ec
\end{figure*}

\subsection{Modified gravity results}
\label{sec:Modified gravity results}
In this section we present the results of the assessment of COLA's  performance with respect to the predicted power spectra, redshift space distortions (RSD) and dark matter halos for the modified gravity scenarios and $\Lambda$CDM. For every given model and choice of parameters, simulations have been performed using both COLA and the PM code.

\subsubsection{Power spectra}
\label{powfr}

To appropriately benchmark the COLA performance for modified gravity, we first compare the performance of COLA for $\Lambda$CDM.  In Fig. \ref{fig3}, we show the $\Lambda$CDM power spectra as obtained by both codes, together in comparison with the fit by \cite{Smith:2002dz}. The two results agree well within a standard deviation of each other for all scales, but start to, underestimate power, consistently with one another, by $k\sim2$ h/Mpc, relative to higher resolution simulations. For that reason, we choose to compare performance down to a scales with $k=2.5$ h/Mpc, while the Nyquist wavenumber, for our simulation, is $k\sim4$ h/Mpc. 

In Fig. \ref{fig4}, the fractional difference in the power spectra is plotted for all our models and both codes are found to agree with each other well within one standard deviation, with the differences being smaller than 1\%. Our results demonstrate the consistency between COLA and the N-body approach using the approximate scheme. In turn this connects with previous work that has shown, in general, the good degree of consistency of this approximate scheme with N-body simulations using the full  Klein Gordon for the same models in the literature \cite{Zhao:2010qy,Davis:2011pj,2012JCAP...10..002B,Winther:2014cia}.   In particular the results for the $\abs{f_{R_0}}=10^{-5}$ \& $\abs{f_{R_0}}=10^{-6}$ models are in excellent agreement with the literature for all scales. Our results confirm findings  in \cite{Winther:2014cia}, in studying the effectiveness of the linearized screening schema: for the lowest screening $f(R)$ model, with $\abs{f_{R_0}}=10^{-4}$, and the symmetron model the effective screening parameterization, respectively, under and overestimates the power, relative to the full KG simulation, at the non-linear scales.  

\begin{figure}[!tb]
\bc
{\includegraphics[width=0.48\textwidth]{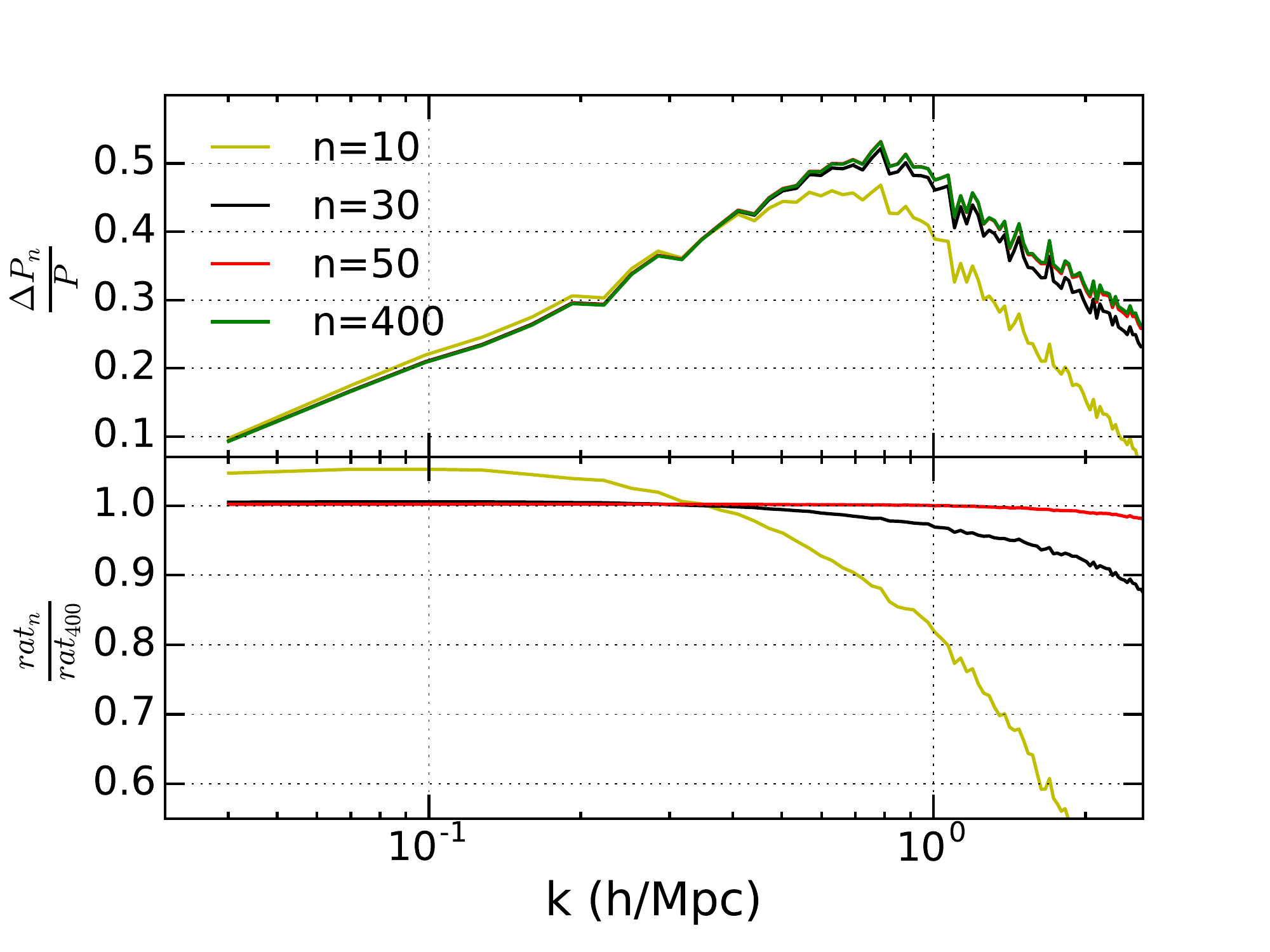}
}
\caption{[Top] Fractional difference in the CDM power spectra for the $\abs{f_{R_0}}=10^{-4}$ scenario relative to $\Lambda$CDM for one realization, as obtained by COLA using various choices of time steps. [Bottom] Ratio of the fractional difference $rat_n=\left(\frac{\Delta P}{P}\right)_n$ for each choice to the high resolution result using 400 steps $rat_{400}$ .}
\label{figlp}
\ec
\end{figure}

Fig. \ref{figlp} shows our COLA scheme's accuracy in predicting the fractional difference in the power spectra for the highest deviation model, $\abs{f_{R_0}}=10^{-4}$,  as a function of the number of time steps used, for one realization. We find that using 50 time steps provides excellent convergence, at the level of 0.9\%, to the scales we want to consider, $k\sim 2$ Mpc/h. Using 30 time steps provides convergence at the level of 8\% at $k\sim 2$ Mpc/h.

\subsubsection{Redshift space distortions}

\begin{figure*}[!tb]
\bc
{\includegraphics[width=0.48\textwidth]{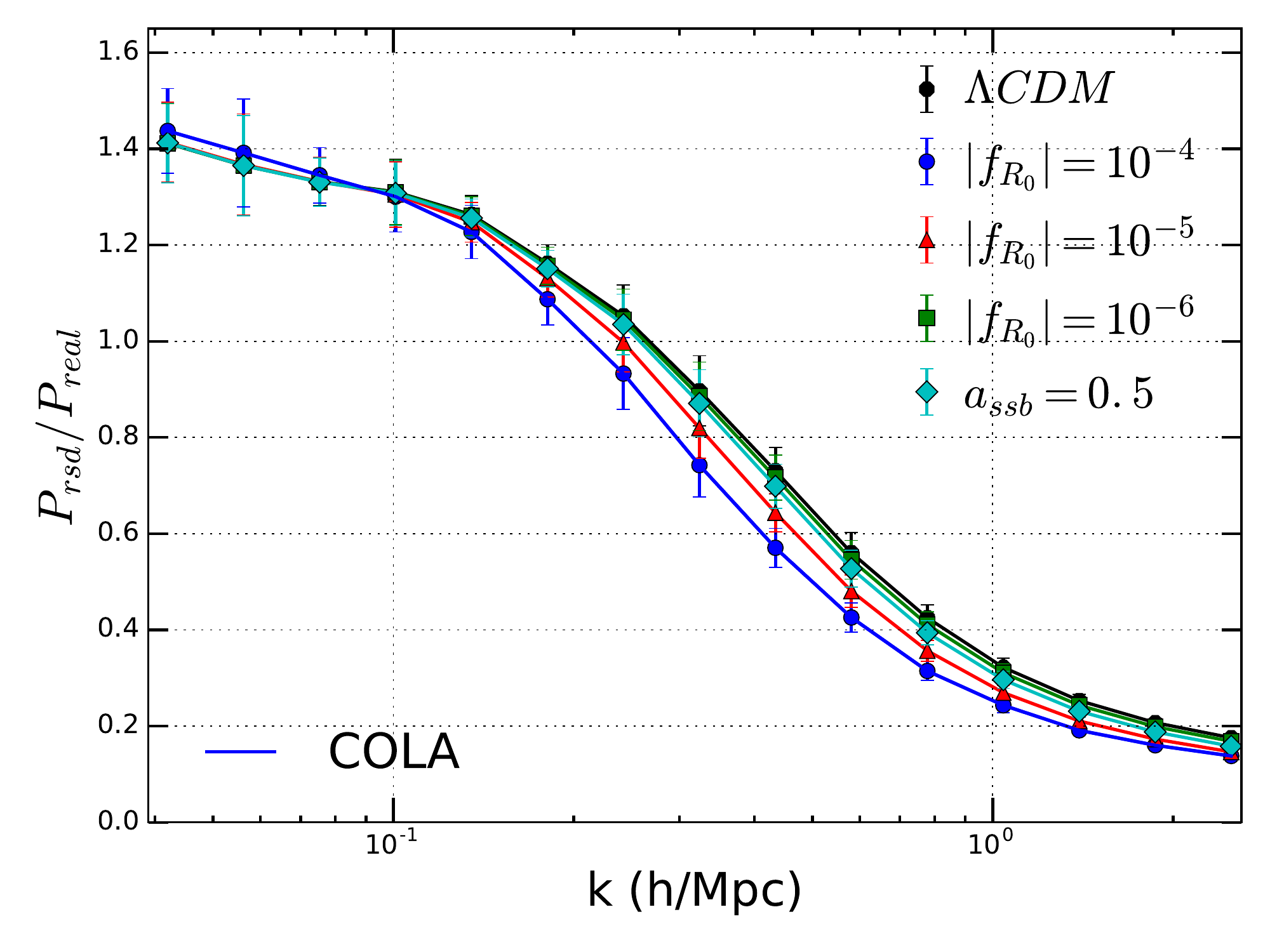}
\includegraphics[width=0.48\textwidth]{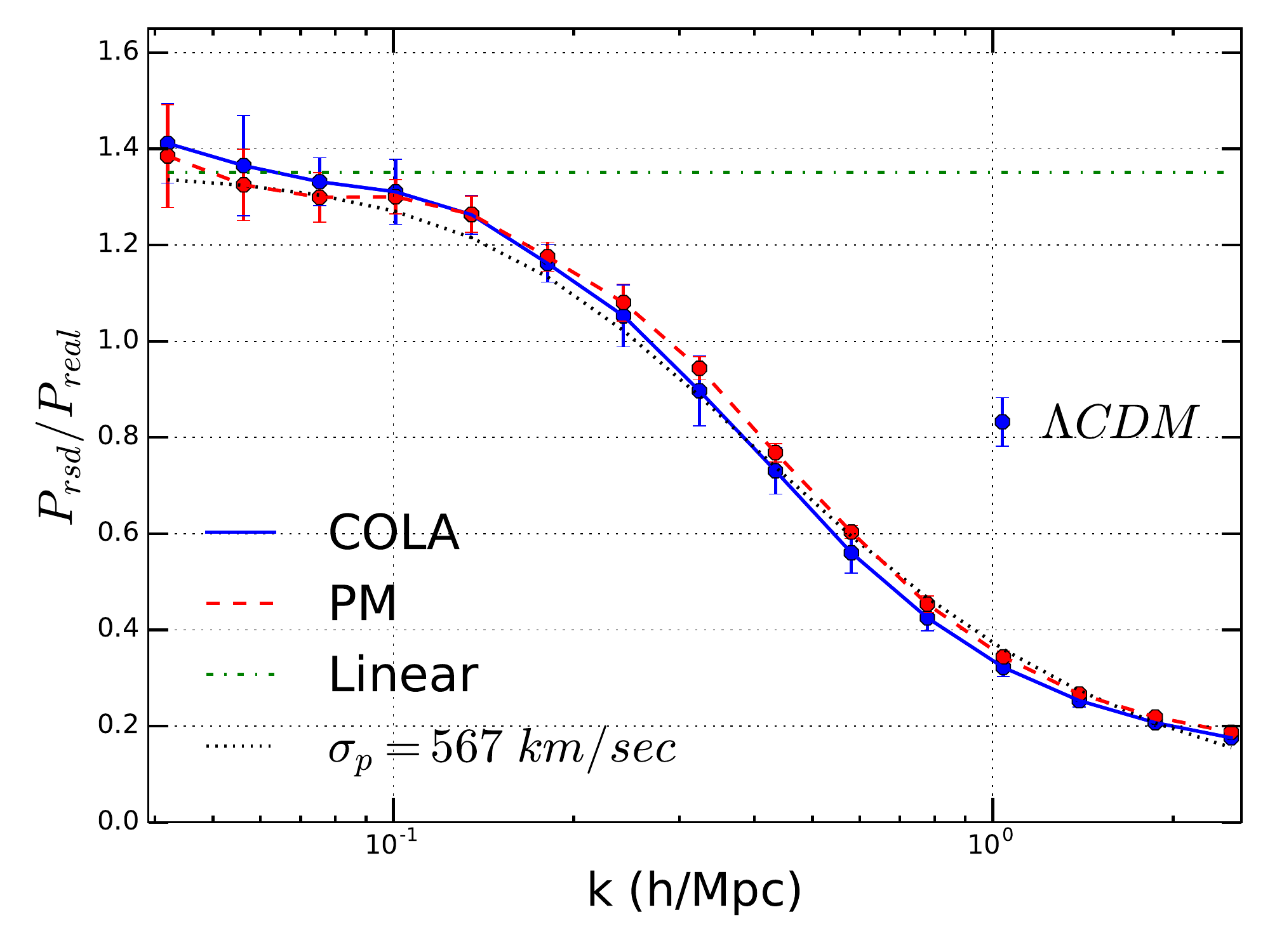}
\includegraphics[width=0.48\textwidth]{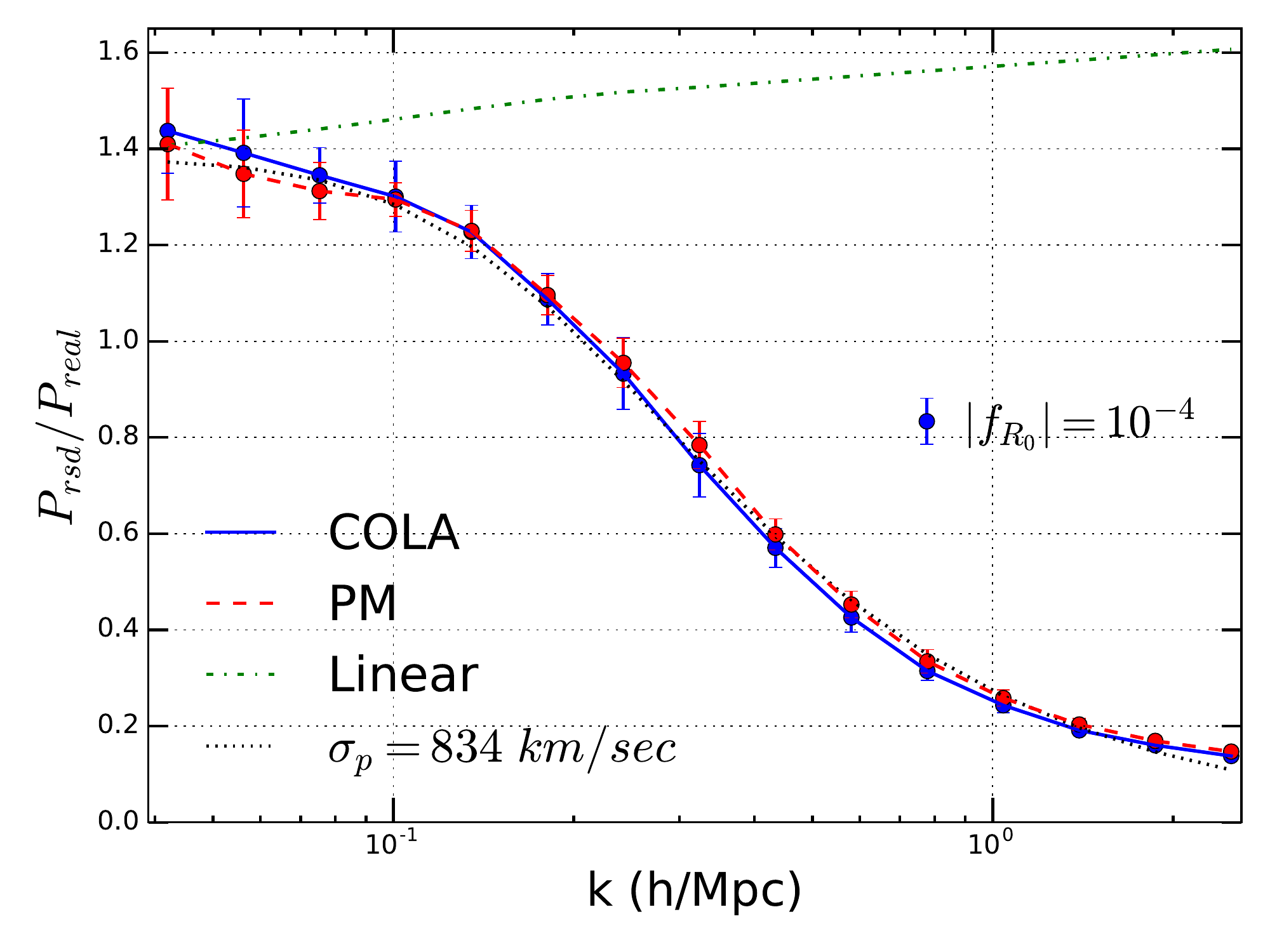}
\includegraphics[width=0.48\textwidth]{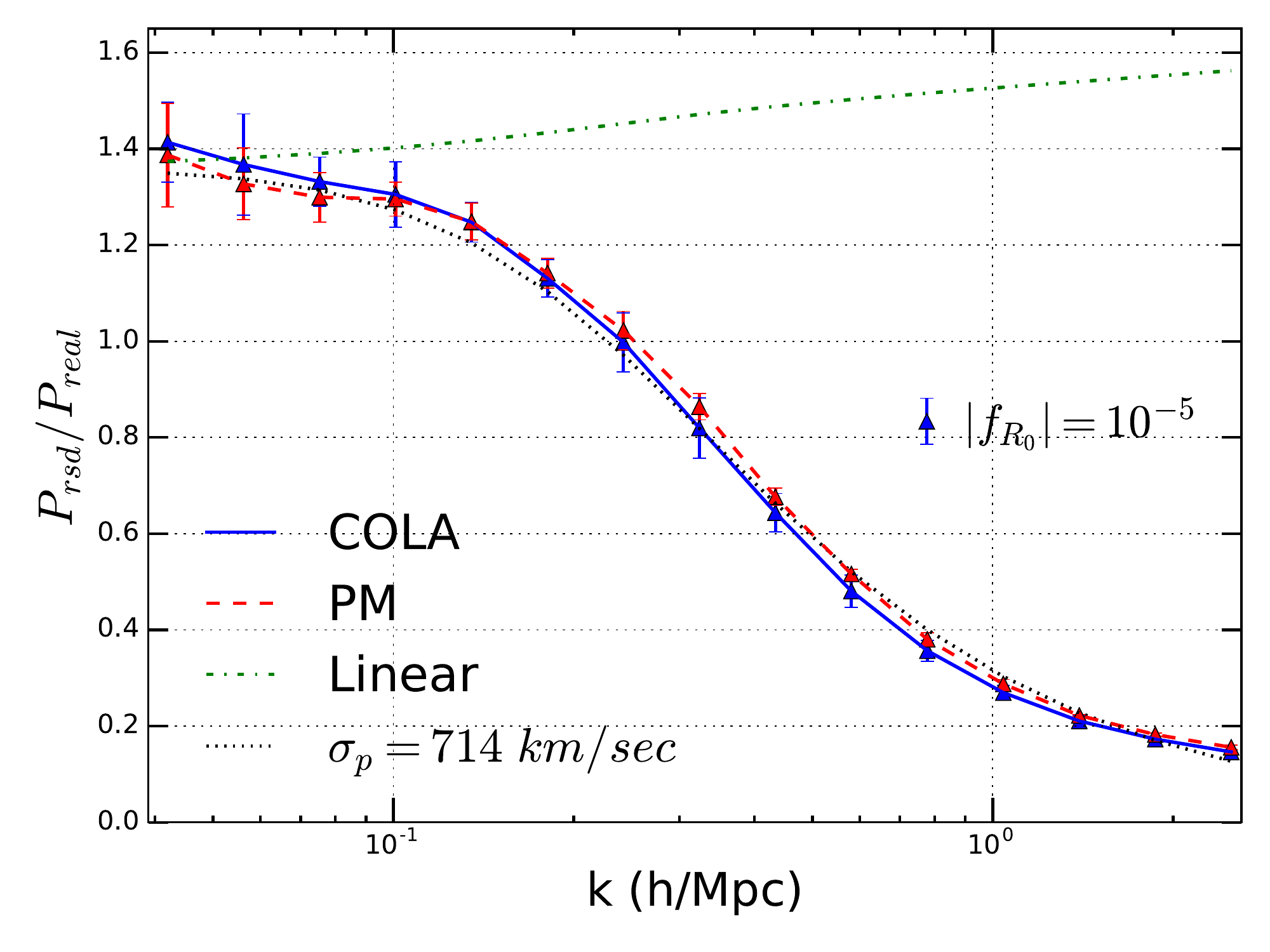}
\includegraphics[width=0.48\textwidth]{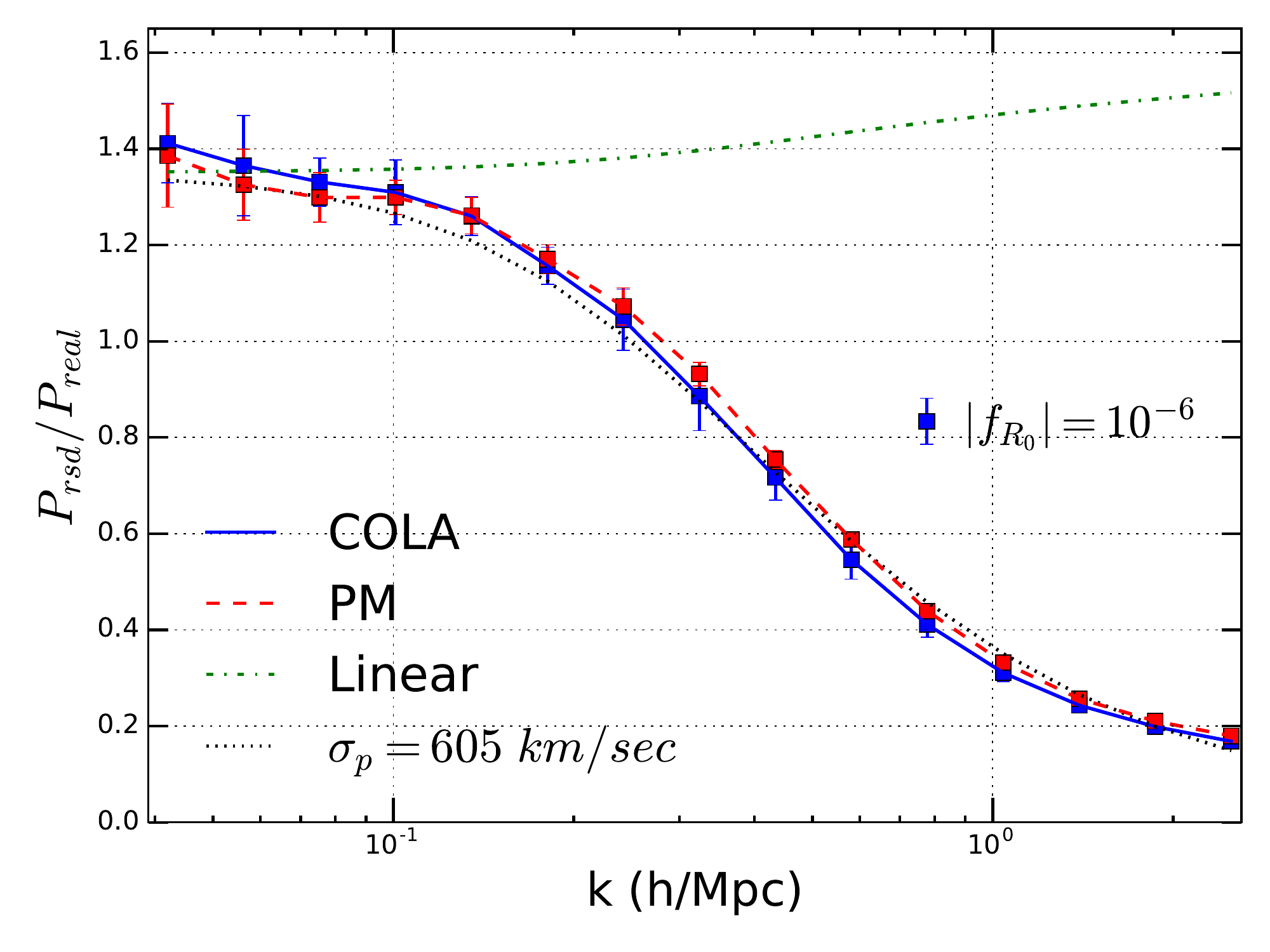}
\includegraphics[width=0.48\textwidth]{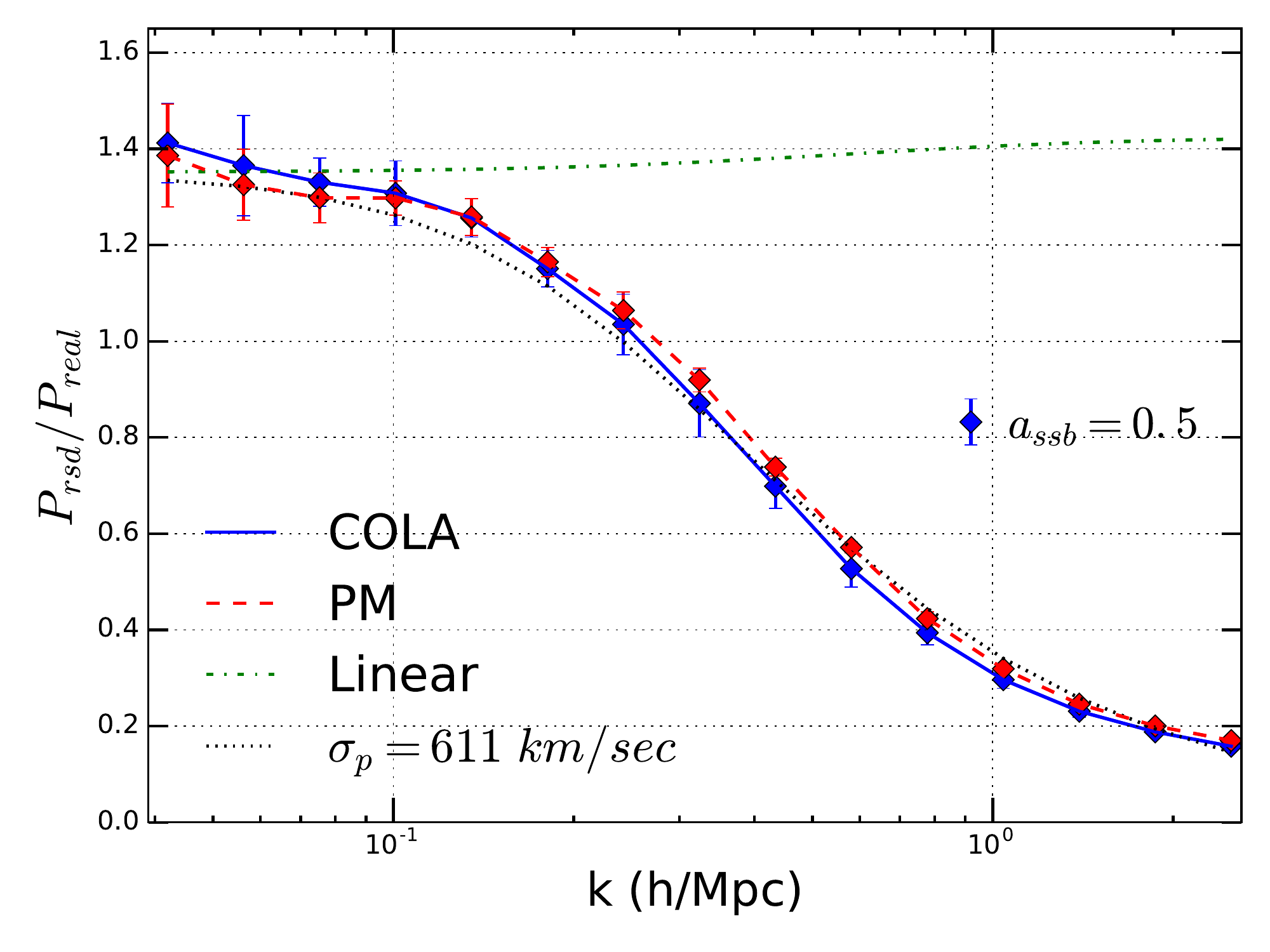}
}
\caption{The ratio of the redshift space power spectrum, $P_{rsd}$, to the real space equivalent, $P_{real}$, for the different models. [Top left] A side-by-side comparison of the suppression of the redshift space clustering by non-linear velocity correlations for the COLA model for $\Lambda$CDM [black], $f(R)$ models with $f_{R_0}=10^{-4},10^{-5}$ and $10^{-6}$ [blue circle, red triangle and green square, respectively], and the symmetron model with $a_{ssb}=0.5$ [cyan diamonds]. The remaining plots show the comparison of PM [red dashed line] and COLA code [blue full] predictions for the RSD to real space power spectrum ratio for each model in turn: [top right] $\Lambda$CDM, [middle left] $f_{R_0}=10^{-4}$, [middle right] $f_{R_0}=10^{-5}$, [bottom left] $f_{R_0}=10^{-6}$, [bottom right] symmetron. Each plot also shows with the linear theory prediction [green dot-dashed] and a fit  to the non-linear suppression using equation (\ref{rsdpow}), and allowing $\sigma_p$ to vary as a free parameter. The number of time steps used for COLA and PM is 50 and 500, respectively.} \label{fig5}
\ec
\end{figure*}

A great amount of observational effort is being invested in studying the three-dimensional Large Scale Structure (LSS) through spectroscopic galaxy surveys that measure precise redshifts. Among various challenges faced by such measurements, the observed clustering structures appear distorted in redshift space. 

Density perturbations give rise to peculiar velocities with respect to the Hubble flow, which result in the redshift space position $\mathbf{r}_s$, being different than the real space position $\mathbf{r}_r$, with the relationship between them taking the form
\begin{equation}
\label{map}
\mathbf{r}_s = \mathbf{r}_r + \frac{\mathbf{v}\cdot\hat{\mathbf{n}}}{H_0}\hat{\mathbf{n}}.
\end{equation}

By $\mathbf{v}$ we denote the peculiar velocity and by $\hat{\mathbf{n}}$ the unit vector along the line of sight. 
At linear scales, coherent motions of galaxies that tend to collapse within an overdense region, cause it to appear squashed in redshift space. As shown by \cite{Kaiser01071987}, in the distant observer approximation, such an overdensity will be distorted in the redshift space:
\begin{equation}
\label{reddel}
\delta_s (\mathbf{k},a) = \left(1 + f \mu^2 \right)\delta_r(\mathbf{k},a),
\end{equation}
where $\mu$ is the angle between the peculiar velocity and the line of sight in $k$ space, $\hat{\mathbf{k}}$, and $f$ the linear growth rate,
\begin{equation}
f_g(a) = \frac{d\ln D_{1}(a)}{d\ln a},
\end{equation}
with subscript `$g$' to differentiate it from the $f(R)$ function. Such an effect gives rise to, based on (\ref{reddel}), an overestimation of the power spectrum measured in the redshift space:
\begin{equation}
\label{redpow}
P_s (k, \mu, a) = \left(1 + \beta_g \mu^2 \right)^2P_r(k,a),
\end{equation}
where we introduced the factor $\beta_g=f_g/b$ (not to be confused with the coupling $\beta$) to account for the galaxy bias $b$, with $b=1$ for cold dark matter. Averaging (\ref{redpow}) over all directions, gives the $0^{th}$ order piece
\begin{equation}
P_s (k,a) = \left(1 + \frac{2}{3}\beta_g + \frac{1}{5}\beta_g^2 \right)P_r(k,a).
\end{equation}
At smaller, non-linear, scales the random incoherent velocities of galaxies within virialized structures cause overdense regions to appear elongated along the line of sight (``Fingers of God"), causing suppression of power. An exact quantitative treatment of the phenomenon is hard, due to the complicated nature of the small-scale velocity correlations and as a result phenomenological approaches have been proposed. Such models  \cite{1992MNRAS.259..494P} treat the line-of-sight distortion as a radial convolution of the correlation function $\xi_{r}$ (including the Kaiser boost) with an incoherent velocity distribution $f(v)$
\be
\xi_s(r_{\perp},r_{\parallel})=\int_{-\infty}^{\infty}\xi_{r}(r_{\perp},r)f(r_{\parallel}-r)dr,
\ee
where $r_{\perp}$ and $r_{\parallel}$ are the perpendicular and parallel components. Assuming a Gaussian velocity distribution \cite{1992MNRAS.259..494P}, the Fourier space expression would then be
\be
\label{gaussf}
P_{s}(k, \mu, a)=P_r(k,a)\left(1 + \beta_g \mu^2 \right)^2\exp(-k^2\mu^2\sigma_{com}^2)
\ee
with $\sigma_{com}$ being the comoving distance dispersion that is related  \cite{2013MNRAS.430.2446W}  to the velocity dispersion $\sigma_{p}$ through
\be
\sigma_{p}=aH(a)\sigma_{com}
\ee
Even though the exponential term in (\ref{gaussf}) is reasonable as a damping term for capturing the non-linear power suppressions, it has been noted \cite{1983ApJ...267..465D} that an exponential pairwise velocity distribution
\be
f(v)=\frac{1}{\sqrt{2}\sigma_p}\exp(-\sqrt{2}\abs{v}/\sigma_p)
\ee
is a  better fit. This gives rise to the dispersion model \cite{Peacock:1993xg}  
\begin{equation}
\label{rsdpow}
P_s (k, \mu, a) = \left(1 + \beta_g \mu^2 \right)^2P_r(k,a)\left(\frac{1}{1 + \frac{1}{2} k^2\mu^2\sigma_{com}^2}\right),
\end{equation}
in which the damping effects are incorporated through a Lorentzian term and $\sigma_{com}$ (or $\sigma_{p}$) is considered a free parameter to be fitted to the data. It should be noted that $\sigma_{p}$ is actually scale and bias dependent, which is one of the limitations the dispersion model faces \cite{Scoccimarro:2004tg}. The above description can still prove to be a very useful tool for obtaining an effective non-linear velocity dispersion parameter and thus quantifying the non-linear FoG effect. Integrating (\ref{rsdpow}) over all directions gives the monopole piece, which can be fitted over the results to obtain $\sigma_{p}$. This is slightly different than other approaches: \cite{Angulo:2007fw} proposed attaching a simple factor $\frac{1}{1+k^2\sigma^2}$ to the Kaiser boost with $\sigma$ being a free parameter, loosely related to $\sigma_{p}$, while \cite{2012ApJ...748...78K} suggested attaching a free function $F(k,\mu)=\frac{A}{1+B k^2\mu^2}+Ck^2\mu^2$ and marginalized over the parameters $A$, $B$ and $C$ for to account for  the uncertainties in constraining the effects of modified gravity on the RSD power spectrum. 
\begin{table}
	\begin{tabular}{ | m{8em}| C{4em} | C{4em} | C{4em} |  C{3.99em} C{0.01em} |}
	\hline 
	& \multicolumn{3}{c|}{Analytic prediction} &  & 
	\\  \cline{2-4}
	 Scenario	 	& ${P_{rsd}/P_{real}}$ 	&   $f_g$ & $\gamma_{eff}$ & $ \sigma_p (km/s)$ &
	\\ [1ex]\hline
		 $\Lambda$CDM  &  	1.35	& 0.462 	& 0.556 &  567	 &
	\\  [1ex]\hline
		$f(R)$, $\abs{f_{R_0}}=10^{-6}$  &1.35  & 0.463 & 0.555 & 605&
	\\ [1ex] \hline
		$f(R)$, $\abs{f_{R_0}}=10^{-5}$  & 1.38 & 0.491& 0.512& 714&
	\\ [1ex] \hline 
		$f(R)$, $\abs{f_{R_0}}=10^{-4}$ &1.42  &  0.541 & 0.443 & 834&
	\\ [1ex] \hline
		Symmetron &1.35 & 0.464 & 0.554 & 611&
	\\ [1ex] \hline
	\end{tabular}
	\caption{Analytic predictions for the RSD to real space power spectrum ratios, $P_{rsd}/P_{real}$,  CDM growth rates, $f_g\equiv dlnD_1(a)/dlna$, and the equivalent growth exponent, $\gamma_{eff}=dln f_g/dln \Omega_m(a)$, are evaluated at $k=0.05 h/Mpc$. The effective velocity dispersion values, $\sigma_p$ are obtained by fitting the average power spectrum results to the monopole of the RSD suppression function in (\ref{rsdpow}).}
	\label{tab1}
\end{table}

Through the mapping (\ref{map}), we obtained redshift space power spectra for all of the simulated models, with the results presented in Fig.\ref{fig5}.  We compared the large scale results to analytic predictions arising from the linear growth rate, and also used the monopole model in (\ref{rsdpow})  to obtain an effective velocity dispersion damping factor for the FoG effect. The results are summarized in  Fig. \ref{fig5} and Table \ref{tab1}. 

We first benchmarked COLA's performance for $\Lambda$CDM. We see that the PM and COLA codes' RSD predictions for $\Lambda$CDM do not differ by more than 0.5\% at all the scales of interest and agree remarkably well with the analytical  prediction, 
with expected values of $f_g(a=1)$=0.467 and $\frac{P_s(k)}{P_r(k)}$=1.354, assuming $f_g=\Omega_m(a)^{\gamma}$, with $\gamma=0.55$. At smaller scales , the ``Fingers of God'' effect  quickly dominates, and causes power suppression and find this suppression is well modeled by (\ref{rsdpow}) with $\sigma_{p}=567\ km/sec$. 

For the MG models, the additional fifth forces cause the redshift space power spectra to have, in principle, different shapes. In large scales, the enhanced clustering results in higher coherent velocities of collapse into overdense regions which translates to a higher boost in the RSD power spectra with respect to GR, translating into higher values for the growth rate and a lower $\gamma$.
For lower magnitude modifications, 
the suppression of the fifth forces gives results 
that tend to the $\Lambda$CDM prediction. At smaller scales, the fifth forces cause higher random velocity dispersions inside virialized structures, making the damping effects stronger in MG. 

\begin{figure}[!tb]
\bc
{\includegraphics[width=0.48\textwidth]{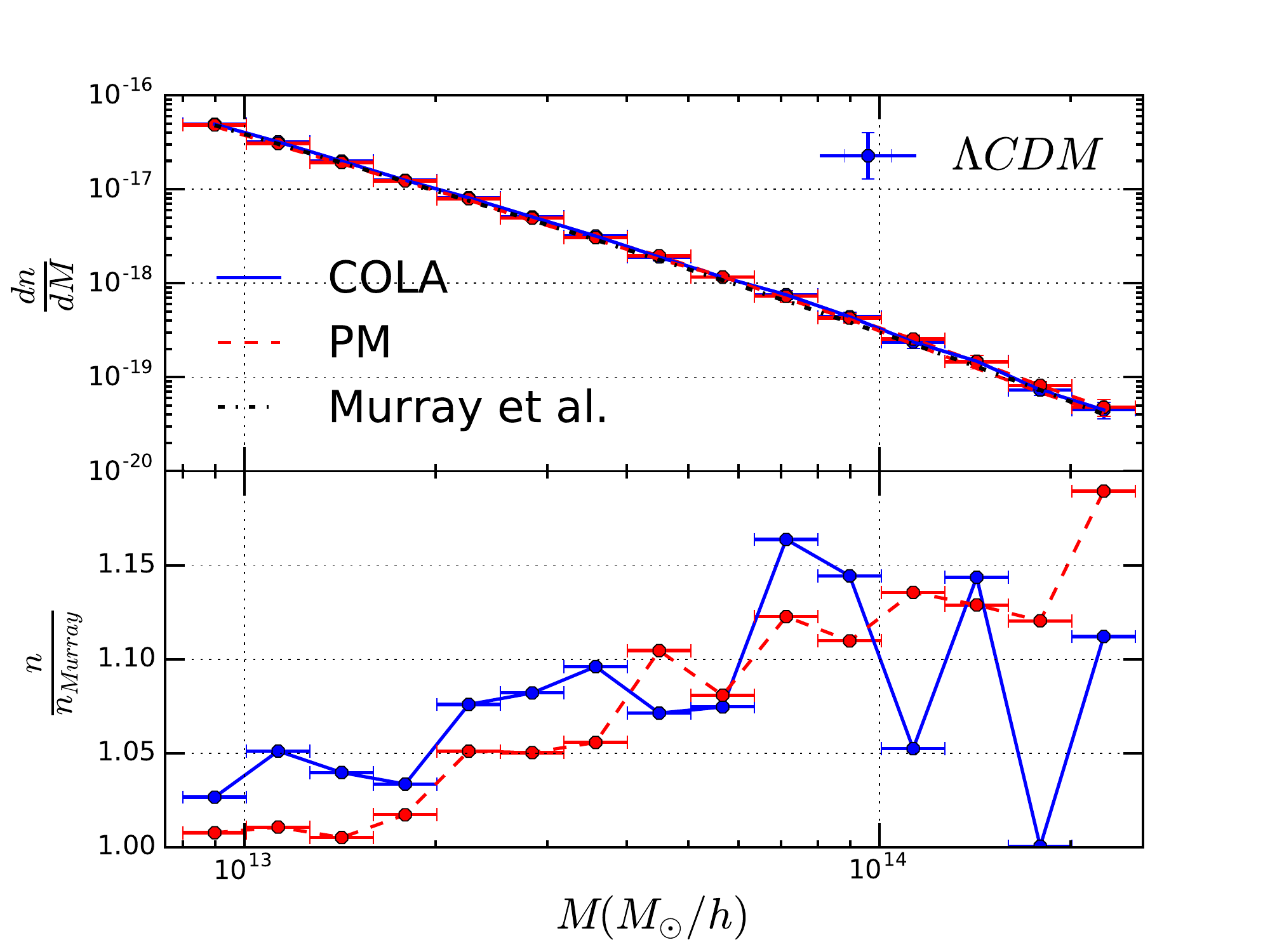}
}
\caption{Halo mass function  benchmarking for $\Lambda CDM$ with the PM N-body code [red dashed line] and COLA method [blue solid line]. The halo mass function fit developed by Murray et al. [black dotted line] is also shown for comparison. The number of time steps used for COLA and PM is 50 and 500, respectively.}
\label{fig6}
\ec
\end{figure}

These combined effects cause the redshift space distortions to be more pronounced in MG compared to GR. This can be clearly seen in the upper left panel in Fig.(\ref{fig5}). As expected, the redshift space distortions vary from the most pronounced, in the lowest screening model, to very small deviations from GR in the strong screening regime. For the same models, the redshift space power spectra from the PM code agree with COLA well within a standard deviation. The results using the approximate schema are in good agreement with full non-linear MG N-body simulations for redshift space distortions in $f(R)$ gravity performed by \cite{2012MNRAS.425.2128J}. For all the models, COLA predicts deviations that are 0.5\% more pronounced (higher in large scales, smaller in small scales) than the PM code.

\begin{figure*}[!tb]
\bc
{\includegraphics[width=0.48\textwidth]{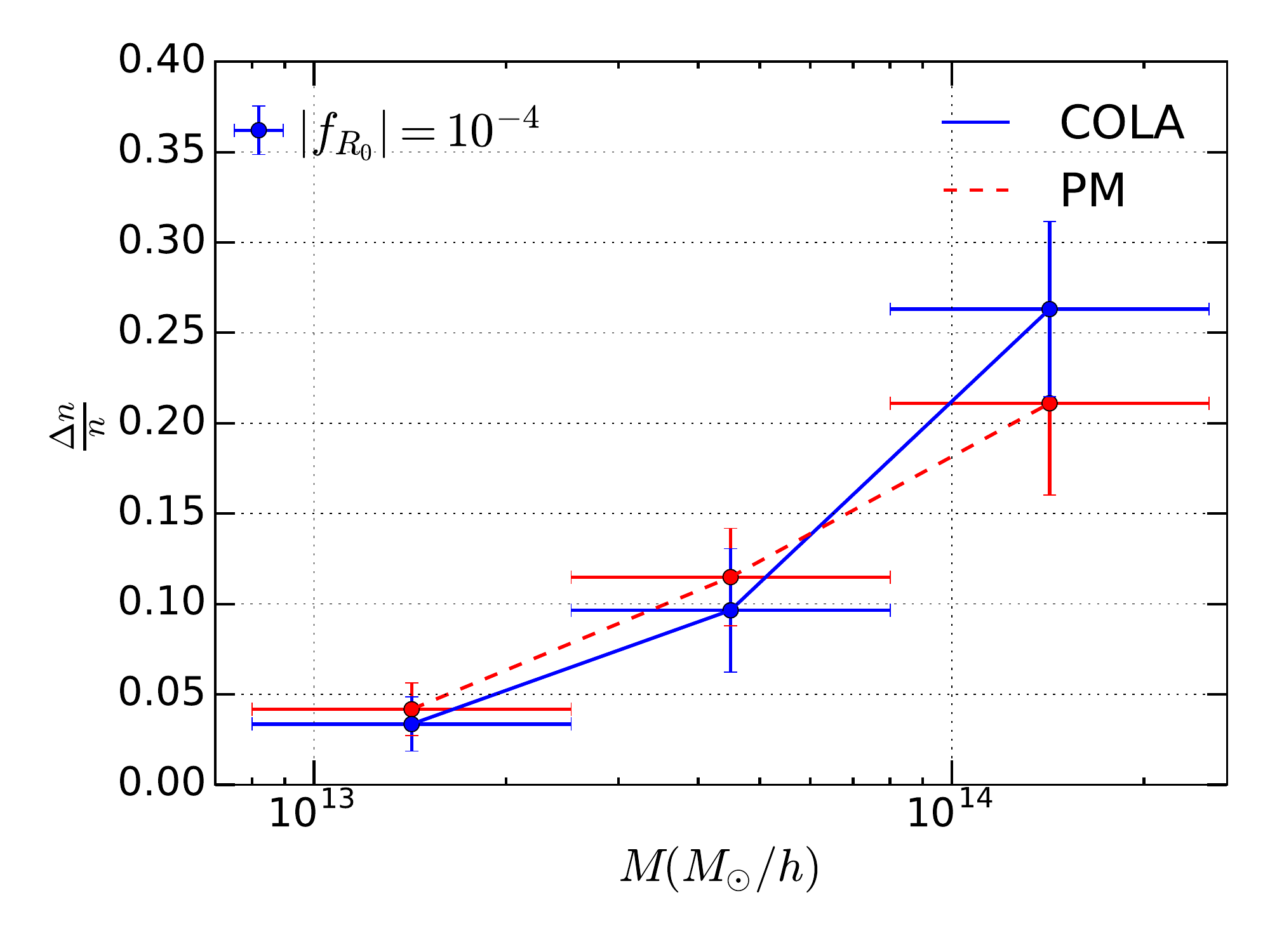}
 \includegraphics[width=0.48\textwidth]{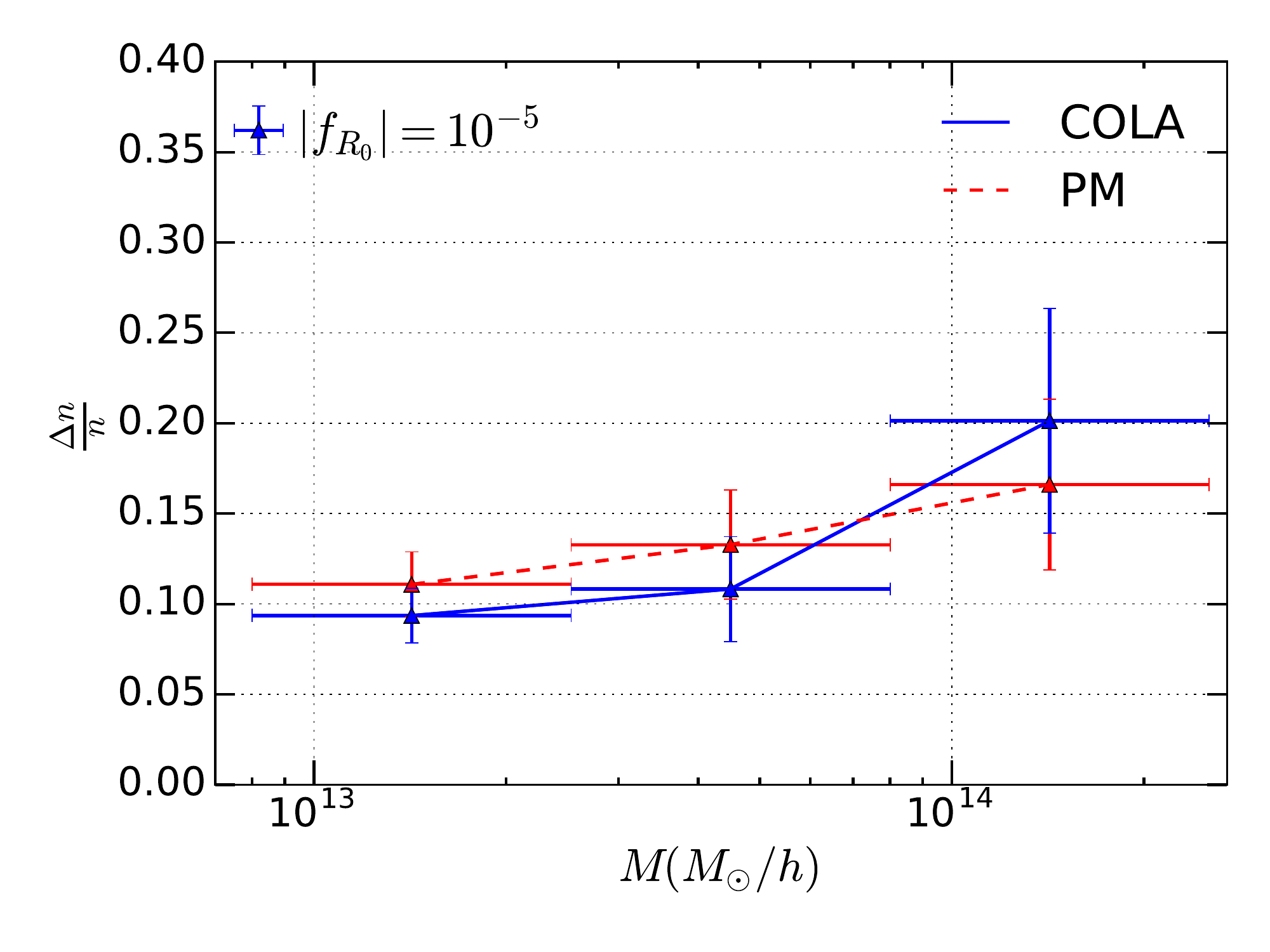}
 \includegraphics[width=0.48\textwidth]{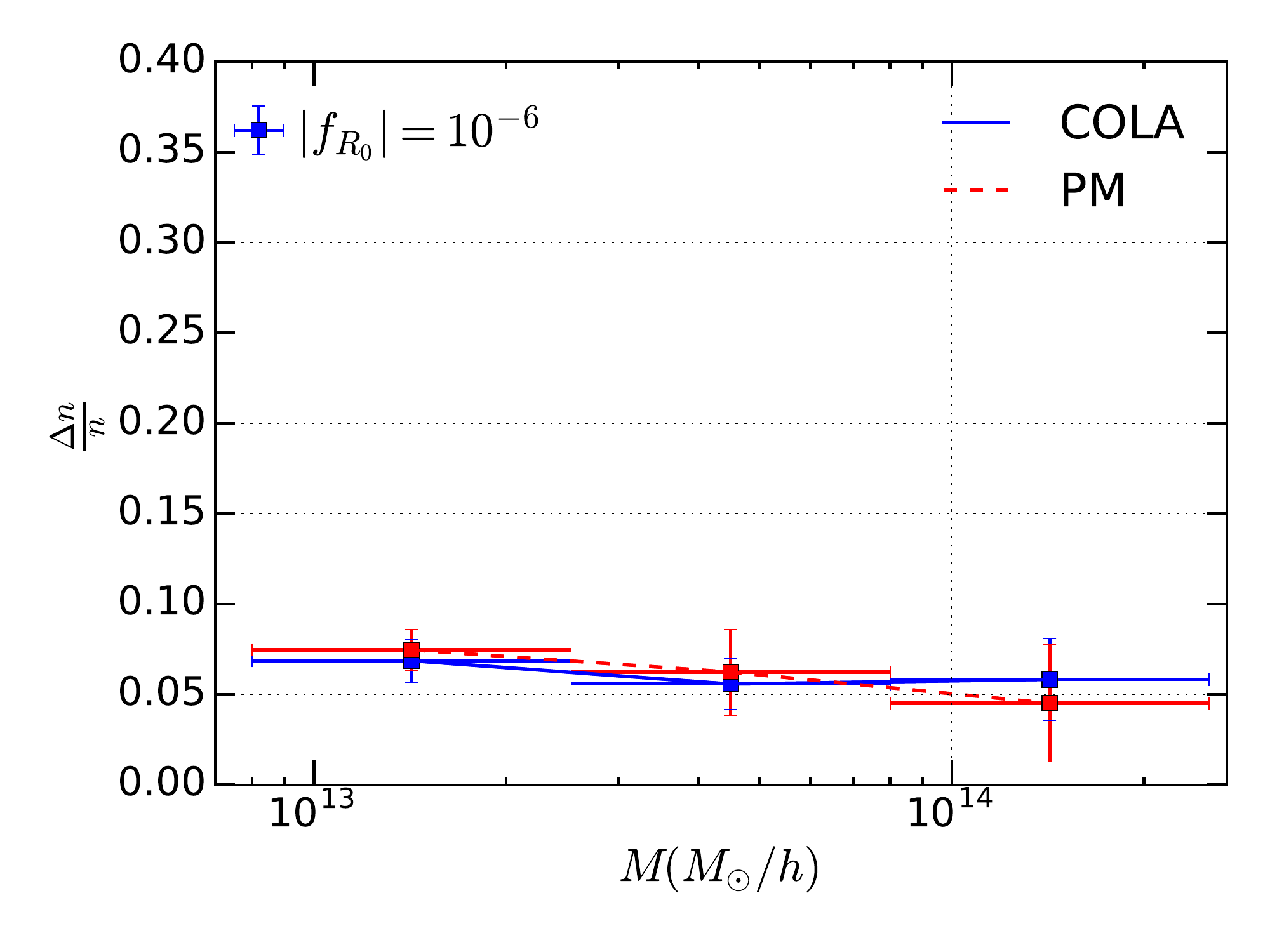}
 \includegraphics[width=0.48\textwidth]{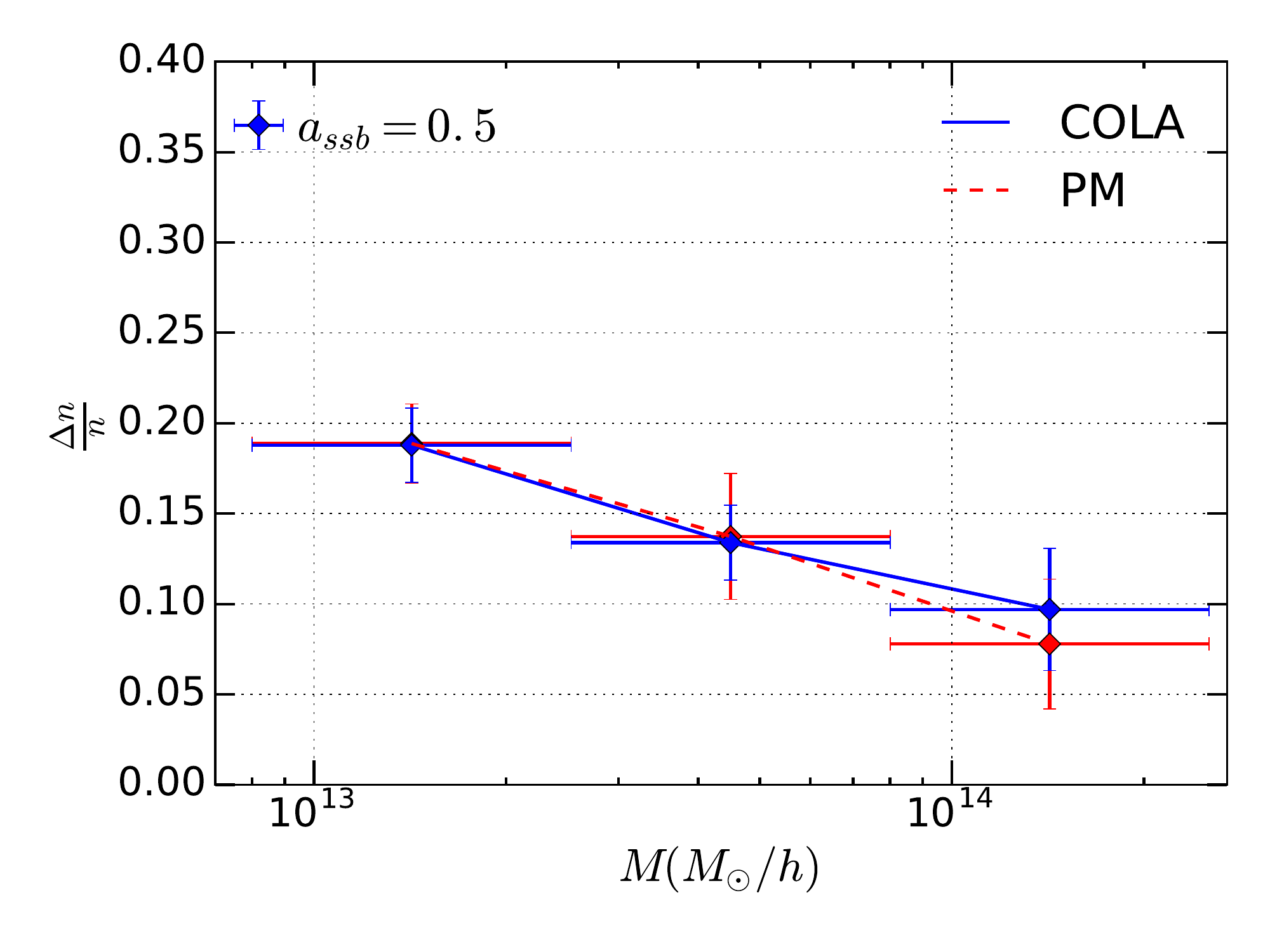}
}
\caption{Fractional difference in the CDM halo mass function, $n(M)$, for halos of mass $M$,  at $a=1$,  for each MG scenario relative to the $\Lambda$CDM model, for the same initial conditions: [left] the $f(R)$ scenario, for $f_{R_0}=10^{-4},10^{-5}$ and $10^{-6}$, and [right]  the symmetron model with $a_{ssb}=0.5$. The averaged results, and standard deviations, from the simulations with the PM code [red dashed line] and the COLA code [blue solid line] are presented in each case. The number of time steps used for COLA and PM is 50 and 500, respectively.}
\label{fig7}
\ec
\end{figure*}

\subsubsection{Halo Mass Function}

To determine the halo mass function we identify halos in the simulations  using the Rockstar halo finder \cite{2013ApJ...762..109B} for all models. In Fig. \ref{fig6} we show the comparison of the halo mass function predicted by COLA and the PM code, together with a high accuracy result by \cite{Murray:2013qza}. COLA and PM are found to be in a better than 2.5\% agreement in the lower and intermediate mass range, while in the highest mass bins there is a maximum difference of 10\%.

In Fig. \ref{fig7}, we plot the fractional difference in the halo mass function with respect to $\Lambda$CDM,  
for all of our models. The COLA and PM code results agree in general, well within the standard deviation from the averaged suite of simulations. In particular, In the $\abs{f_{R_0}}=10^{-4}$ and $\abs{f_{R_0}}=10^{-5}$ models, the PM code predicts a fractional boost in the halo mass function that is higher than COLA's by $<$ 2\% and 2.5\%, for the lower and intermediate bins, while in the highest bin COLA gives a boost larger by 5\% and 3\% correspondingly. For the $\abs{f_{R_0}}=10^{-6}$ and symmetron models, the differences are 1\% and smaller, with the PM code giving greater number counts 
for the two mass bins below $10^{14}M_{\odot}/h$ and COLA being higher for the bin over $10^{14}M_{\odot}/h$.
The differences between the predictions in each case and especially in the high mass bin, are within, and likely largely resulting from, the differences observed in the $\Lambda$CDM benchmarking of the mass functions. 
 
While we do not perform a simulation with the full non-linear Klein-Gordon equation, we note that compared to other full KG treatments in the literature \cite{Winther:2014cia,Zhao:2010qy}, our method performs well and only slightly underestimates the mass function for the $\abs{f_{R_0}}=10^{-4}$ \& $\abs{f_{R_0}}=10^{-5}$ models, in accordance with the general features noticed in the power spectra discussion. In agreement with  \cite{Winther:2014cia}, we observe an underestimation of halos in the lower end of our mass range (around M $\sim10^{13}{M_{\odot}/h}$) for the $\abs{f_{R_0}}=10^{-6}$ model, indicating too much screening, and an overestimation of the mass function for the symmetron model.

\section{Conclusions }
\label{sec:conclusions}
In this paper, we have implemented a hybrid scheme, that combines Lagrangian Perturbation Theory and N-body approaches, to numerically  characterize the evolution of large scale structure in chameleon and symmetron, modified gravity, theories which exhibit gravitational screening in the non-linear regime. LPT is used  to evolve linear scales analytically in combination with a full N-body approach that is used for the non-linear scales to reduce computational costs. An effective screening scheme is implemented in place of a solution to the full Klein-Gordon equation for the fifth potential, in which an effective suppression factor is attached to the real-space linearized perturbations. 

We demonstrate that while in MG spatial modes evolve differently in LPT (and can have deviations from the nominal GR geodesic paths), the scheme can be further simplified,  for the models we studied, by using a displacement coordinate system based on scale-independent $\Lambda$CDM growing modes   combined with a modified, screened Poisson equation. We note that, while this approximate scheme works well for the chameleon and symmetron models we consider, it should always be tested against the exact LPT solution for a new modified gravity model.
 
Our method was applied on the $f(R)$ and symmetron models and it was tested against power spectra, redshift space distortions and dark matter halo mass functions, using a fiducial number of 50 time steps. At the same time, we assessed our hybrid's performance against simulations from a pure N-body code with the same screening implementation for the same models, using 500 iterations. 

With regards to power spectra, we found COLA to be in better than 1\% agreement with the N-body code at all scales for all the models studied.
Note that the effective screening scheme we use has previously been shown to be in good agreement with results using the full non-linear Klein Gordon  in an N-body implementation \cite{Winther:2014cia}. We find, as was discussed in \citep{Winther:2014cia}, that the effective screening approach  does underestimate power, relative to that found in solving the full Klein-Gordon \cite{Zhao:2010qy}, as one moves into the fully non-linear regime ($k>\sim 2Mpc/h$),  however this is also beyond the regime of applicability of COLA's scheme. 

COLA and the N-body code are in better than 0.5\% agreement with respect to redshift space distortions for all the scales and models of interest. The distortions were modeled by attaching the linear Kaiser factor for the enhancement at large scales and a Lorentzian dispersion factor for the small scale suppression due to incoherent motions within virialized structures. We find that the monopole is a well fit using an effective pairwise velocity dispersion as a fitting parameter to quantify the suppressions at non-linear scales. The additional fifth forces present in the chameleon and symmetron models, cause the redshift space distortions to be more pronounced with respect to $\Lambda$CDM. This can be seen by the larger boosts in linear scales due to the higher coherent velocities, and by the stronger suppressions in the non-linear scales because of the higher values of the velocity dispersion. The adapted COLA scheme gives reasonable results for the predicted fractional boost in the halo mass function relative to $\Lambda$CDM, with the differences between the N-body and COLA results in the halo mass function estimation most likely being due to the difference between the two codes in $\Lambda$CDM. 

In this paper, we have focused  on chameleon and symmetron-type scalar-tensor theories, but it would be very interesting to see how well this scheme performs for the simulation of other screening mechanisms as well such as the Vainshtein mechanism \citep{VAINSHTEIN1972393}, as well as other dark energy models, such as those with non-minimal couplings between dark matter and a quintessence scalar field \citep{Li:2010re}. 
Given the level of consistency between COLA and the N-body predictions for the monopole of the redshift power spectrum, it would also be interesting to  investigate the COLA scheme's ability to capture higher order moments of the angular power spectrum to, for example, calculate the ratio of the quadrupole to monopole moments to estimate $\beta_g$ in a way that is robust to systematic effects from incomplete modeling of the nonlinear distortions \cite{Landy:2002eq,2012MNRAS.425.2128J}.

Many theories being considered as  explanations for cosmic acceleration have tantalizing predictions in the non-linear regime but also present computational challenges in modeling them.  With a suite of next-generation  large scale structure surveys, including LSST, DESI, Euclid and WFIRST, starting in next few years, there is an unprecedented opportunity to measure the properties of large scale structure clustering as it transitions from linear to mildly and then strongly non-linear scales, and using multiple tracers.  The results presented here demonstrate that COLA, proposed to enable accurate and efficient, non-linear predictions for $\Lambda$CDM, is a viable  approach to study non-linear collapse for a broader portfolio of cosmological scenarios. For example, in work that has followed our paper in \cite{Winther:2017jof}, the effectiveness of the COLA approach has also been studied in the $f(R)$ and nDGP models, and was shown to perform very well in predicting the fractional deviations with respect to the $\Lambda$CDM power spectra and halo mass functions, using a small number of time steps.

\section*{Acknowledgments}

We want to thank Hans Winther and Pedro Ferreira for useful comments on our paper and Risa Wechsler for helpful discussions on halo mass function performance with the COLA algorithm. We would also like to thank an anonymous referee, whose valuable comments helped improve and clarify this manuscript. The work of GV and RB is supported by NASA ATP grant NNX14AH53G, NASA ROSES grant 12-EUCLID12- 0004 and DoE grant DE-DE-SC0011838.

\appendix
\section{\\Lagrangian perturbation theory in Modified Gravity} 
\label{App:AppendixA}
LPT \cite{Bouchet:1994xp} works perturbatively in a displacement field $\mathbf{s}(\mathbf{q},a)$
\begin{equation}
\mathbf{x}=\mathbf{q}+\mathbf{s}_{1}(\mathbf{q},a)+\mathbf{s}_{2}(\mathbf{q},a)+...,
\end{equation}
where $\mathbf{q}$ and $\mathbf{x}$ are the initial and final comoving Eulerian particle positions. In this formulation, all the information is reflected in the mapping through the displacement field. Working up to first order gives the so-called Zel'dovich approximation in $\Lambda$CDM, for which the $\mathbf{s}(\mathbf{q},a)$ can be decomposed into a product of temporal and spatial factors
\begin{equation}
\nabla_{\mathbf{q}}\mathbf{s}_{1}(\mathbf{q},a) = D_{1}(a)\nabla_{\mathbf{q}}\mathbf{s}_{1}(\mathbf{q},a_0)
\end{equation}
and
\begin{equation}
\label{gauss}
\nabla_{\mathbf{q}}\mathbf{s}(\mathbf{q},a_0) = -\delta(\mathbf{q},a_0)
\end{equation}
with $\delta(\mathbf{q},a_0)$ being the Gaussian density field generated by an initial linear power spectrum and $D_{1}(a)$ the scale independent first order growth factor, given by
\begin{equation}
\ddot{D}_{1} + 2H\dot{D}_{1} = \frac{3}{2} \Omega_m(a)H^2 D_{1} 
\end{equation}
In an MG scenario, the growth factor is not scale independent any more and
\begin{equation}
\label{firstdisp}
\nabla_{\mathbf{q}}\mathbf{s}_{1} = D_{1}(\mathbf{q},a)\nabla_{\mathbf{q}}\mathbf{s}_1(\mathbf{q},a_0)
\end{equation}
where
\begin{equation}
\ddot{D}_{1}(\mathbf{k},a) + 2H\dot{D}_{1}(\mathbf{k},a) = \frac{3}{2} \Omega_m(a)H^2 D_{1}(\mathbf{k},a)\frac{G_{eff}}{G}
\end{equation}
in Fourier space.
This implies that particle trajectories, unlike in $\Lambda$CDM, are not straight lines \cite{Valkenburg:2015dsa}. (\ref{gauss}) and (\ref{firstdisp}) indeed give 
\begin{equation}
\label{fullZel}
\mathbf{s}_{1} = -D_{1}(\mathbf{q},a)\frac{\nabla_{\mathbf{q}}}{\nabla^2_{\mathbf{q}}}\delta(\mathbf{q},a_0)-\delta(\mathbf{q},a_0)\frac{\nabla_{\mathbf{q}}}{\nabla^2_{\mathbf{q}}}D_{1}(\mathbf{q},a)
\end{equation}
The second term, responsible for the trajectory bending, vanishes when the growing mode is scale independent, in which case one recovers the standard $\Lambda$CDM Zel'dovich approximation.
When working up to second order (2LPT), we have in a similar fashion
\begin{equation}
\nabla_{\mathbf{q}}\mathbf{s}_{2}(\mathbf{q},a) = D_{2}(a)\nabla_{\mathbf{q}}\mathbf{s}_{2}(\mathbf{q},a_0),
\end{equation}
where the second order growth factor is given by
\begin{equation}
\ddot{D}_{2}(a) + 2H\dot{D}_{2}(a) = \frac{3}{2} \Omega_m(a)H^2 D_{2}(a)\left(1 - D_1^2(a)\right)
\end{equation}
For the early times, the spatial part is given by
\begin{equation}
\label{spatial}
\nabla_{\mathbf{q}}\mathbf{s}_{2}(\mathbf{q},a_0) = \frac{1}{2}\sum_{i\ne j}\left(\mathbf{s}_{1i,i}\mathbf{s}_{1j,j}-\mathbf{s}_{1i,j}\mathbf{s}_{1j,i}\right).
\end{equation}
In the MG case, we will have again
\begin{equation}
\label{seconddisp}
\nabla_{\mathbf{q}}\mathbf{s}_{2}(\mathbf{q},a) = D_{2}(\mathbf{q},a)\nabla_{\mathbf{q}}\mathbf{s}_{2}(\mathbf{q},a_0),
\end{equation}
with the scale dependent second order growth factor that obeys
\begin{equation}
\begin{split}
\ddot{D}_{2}(\mathbf{k},a) + 2H\dot{D}_{2}(\mathbf{k},a) & = \frac{3}{2} \Omega_m(a)H^2 D_{2}(\mathbf{k},a)\times \\
                               						          &    (1 - D_1^2(\mathbf{k},a))\frac{G_{eff}}{G},
\end{split}
\end{equation}
in the Fourier space.
The fact that all of our models recover GR at early times, guarantees that the early time spatial part is still given by (\ref{spatial}). In our implementation of the full MG COLA scheme, a suitably modified version of 2LPTic produces the LPT terms at every time step, through the Fourier space versions of (\ref{firstdisp}) and (\ref{seconddisp})
\begin{equation}
\label{fourdisps}
\begin{split}
\mathbf{s}_{1}(\mathbf{k},a) & = \frac{i \mathbf{k}}{k^2}D_{1}(\mathbf{k},a)\delta(\mathbf{k},a_0) \\
\mathbf{s}_{2}(\mathbf{k},a) & = -\frac{i \mathbf{k}}{k^2}D_{2}(\mathbf{k},a) \frac{1}{2}\sum_{i\ne j}\left(\mathbf{s}_{1i,i}\mathbf{s}_{1j,j}-\mathbf{s}_{1i,j}\mathbf{s}_{1j,i}\right),
\end{split}
\end{equation} 
and also the same for the accelerations
\begin{equation}
\label{fouraccels}
\begin{split}
T^2[\mathbf{s}_{1}(\mathbf{k},a)] & = \frac{i \mathbf{k}}{k^2}T^2[D_{1}(\mathbf{k},a)]\delta(\mathbf{k},a_0) \\
T^2[\mathbf{s}_{2}(\mathbf{k},a)] & = -\frac{i \mathbf{k}}{k^2}T^2[D_{2}(\mathbf{k},a)] \frac{1}{2}\sum_{i\ne j}\left(\mathbf{s}_{1i,i}\mathbf{s}_{1j,j}-\mathbf{s}_{1i,j}\mathbf{s}_{1j,i}\right).
\end{split}
\end{equation} 
\newline
\newpage
\nocite{*}
\bibliographystyle{apsrev}

\end{document}